%% file: main.tex
\documentclass[journal]{IEEEtran}
\renewcommand\IEEEkeywordsname{Index Terms}
\usepackage{blindtext}
\usepackage{gensymb}
\usepackage{graphicx}
\usepackage{amssymb}
\usepackage{amsmath}
\usepackage{amsfonts}
\usepackage{xcolor}
\usepackage{braket}
\usepackage{multirow}
\usepackage{subfigure}
\usepackage{cite}
\usepackage[ruled]{algorithm2e}
\usepackage[utf8]{inputenc}
\usepackage[english]{babel}
\usepackage{caption}
\usepackage{hyperref}
 
\usepackage{amsthm}
\ifCLASSINFOpdf
\else
\fi
\input{newcommands}

\begin{document}
%
\newtheorem{corol}[theorem]{Corollary}
\title{Target Tracking With ISAC Using EMLSR in Next-Generation IEEE 802.11 WLANs: Non-Cooperative and Cooperative Approaches}

%
%
%

\author{
Ching-Lun Tai,~\IEEEmembership{Graduate Student Member,~IEEE,} Jingyuan Zhang,~\IEEEmembership{Graduate Student Member,~IEEE,}\\ Douglas M. Blough,~\IEEEmembership{Senior Member,~IEEE,} and~Raghupathy Sivakumar,~\IEEEmembership{Fellow,~IEEE}
\thanks{This work was supported by the Wayne J. Holman
Chair and the EVP for Research at Georgia Tech. The authors are with the School of Electrical and Computer Engineering, Georgia Institute of Technology, Atlanta, GA 30332 USA (e-mail: ctai32@gatech.edu; jingyuan\_z@gatech.edu; doug.blough@ece.gatech.edu; siva@ece.gatech.edu).}
}

\maketitle
\thispagestyle{empty}

\begin{abstract}
New amendments support Wi-Fi access points (APs) and stations (STAs) in next-generation IEEE 802.11 wireless local area networks (WLANs).
IEEE 802.11be (Wi-Fi 7) features multi-link operation (MLO) with multi-link device (MLD) hosting multiple interfaces, highlighting enhanced multi-link single-radio (EMLSR) operation.
IEEE 802.11bf features Wi-Fi sensing, enabling integrated sensing and communications (ISAC) in Wi-Fi.
In this paper, we pioneer an innovative combination of EMLSR operation and ISAC functionality, considering target tracking with ISAC using EMLSR in IEEE 802.11 WLANs.
We establish a unique scenario where AP MLD needs to make ISAC decision and STA MLD selection when its interface gains a transmit opportunity (TXOP).
Then, we present key design principles: ISAC decision involves the Kalman filter for target state and a developed time-based strategy for sensing/communications determination, while STA MLD selection involves a Cram{\'e}r-Rao lower bound (CRLB)-based trilateration performance metric along with a developed candidate strategy for UL sensing and involves a developed weighted proportional fairness-aware heuristic strategy for DL communications.
We propose novel non-cooperative and cooperative approaches, where each interface leverages its own information and aggregate information across all interfaces, respectively.
For proposed non-cooperative and cooperative approaches, simulation results exhibit their tradeoff and superiority about sensing and communications.

\end{abstract}


%
\IEEEpeerreviewmaketitle

\section{Introduction}


\IEEEPARstart{F}{or} more than two decades, Wi-Fi has been the most popular wireless local area network (WLAN) technology broadly deployed all over the world, and the global economic value of Wi-Fi is envisioned to reach around 5 trillion USD by 2025 \cite{wfa2021}.
Due to the emergence of various applications with more stringent requirements and more diverse needs, new amendments have been proposed to provide essential support for next-generation IEEE 802.11 WLANs.

To achieve faster data transmission in the presence of an ever-increasing traffic demand, the IEEE 802.11be extremely high throughput (EHT) amendment (commercially known as Wi-Fi 7) launches several features.
One of the key features launched in IEEE 802.11be is multi-link operation (MLO) \cite{lopez2022b}, which introduces the architecture of multi-link device (MLD) that enables an operation over multiple links for both access point (AP) and station (STA).
Particularly, an AP MLD or an STA MLD manages multiple links by hosting multiple interfaces, each of which has its own link(s) and transmit opportunity (TXOP) over the channel at a specific frequency band.
Prior research has investigated the coexistence of legacy single-link devices (SLDs) and MLDs (e.g., \cite{murti2021,korolev2021,medda2022,korolev2022}) and have evaluated the performance of MLO under different scenarios (e.g., \cite{zhang2022,carrascosa2022,alsakati2023,bellalta2023,carrascosa2023,carrascosa2024,kulshrestha2024}).

Among a variety of MLO types, the enhanced multi-link single-radio (EMLSR) operation is the most promising option, requiring less hardware installation and lower implementation cost \cite{adhikari2022}.
Under an EMLSR operation, an interface of the AP MLD gains its own TXOP and interacts with the corresponding interface of some STA MLDs which are listening to their links and awaiting subsequent frame exchange.
A couple of previous works have studied the dynamic channel allocation for EMLSR operation with different schemes.
For instance, \cite{tai2024b} proposes a fairness-aware heuristic method, and \cite{chen2024} proposes a neural network (NN)-based mechanism.

In addition to communications, sensing (for environment observation) is expected to become another main task of Wi-Fi due to its widespread ubiquity \cite{tan2022}.
Thus, the IEEE 802.11bf amendment officially launches the feature of Wi-Fi sensing \cite{chen2023}, where an AP or an STA (including an interface of AP MLD or STA MLD) serves as a sensing initiator (SI) or a sensing responder (SR).
For a sensing activity, the SI initiates a sensing procedure with the SR involved.
A detailed description of the sensing procedure for Wi-Fi sensing in IEEE 802.11bf can be found in \cite{ropitault2024}.

Taking advantage of both communications paradigm and sensing capability, integrated sensing and communications has been an active area of research \cite{lu2024}.
With its well-developed infrastructure for communications and sensing, Wi-Fi can significantly benefit from ISAC \cite{meneghello2023}.
Some prior research has studied the adoption of ISAC in Wi-Fi.
For example, \cite{tai2024c} proposes a new method for target tracking with ISAC in Wi-Fi, and \cite{he2024} builds a prototype accommodating ISAC within commercial Wi-Fi devices.

Toward IEEE 802.11 WLANs, the sensing and communications performance of ISAC can be improved with the use of EMLSR.
However, an innovative combination of EMLSR operation (supported by IEEE 802.11be) and ISAC functionality (supported by IEEE 802.11bf) has been less explored in the context of Wi-Fi within the existing literature.

Therefore, in this paper, we focus on target tracking with ISAC using EMLSR in IEEE 802.11 WLANs.
Particularly, we establish a unique scenario, present key design principles, and propose novel non-cooperative and cooperative approaches.
The following summarizes our key contributions:
\begin{itemize}
    \item We construct distinctive system model and problem formulation for target tracking with ISAC using EMLSR in IEEE 802.11 WLANs, where the AP MLD needs to make an ISAC decision and an STA MLD selection when its interface gains a TXOP. 
    \item Then, we present the key design principles of ISAC decision and STA MLD selection. The ISAC decision tracks the state of target with the Kalman filter \cite{kim2018introduction} and determines an action between sensing and communications with a developed time-based strategy. The STA MLD selection takes a Cram{\'e}r-Rao lower bound (CRLB) \cite{nielsen2013cramer}-based trilateration performance metric along with a developed candidate strategy for UL sensing and takes a developed weighted proportional fairness \cite{shi2014}-aware heuristic strategy for DL communications.
    \item According to the design principles, we propose a non-cooperative approach, where each interface of the AP MLD employs its own information upon gaining a TXOP under an EMLSR operation. With regard to the proposed non-cooperative approach, we describe its procedures of ISAC decision and STA MLD selection, and we analyze its computational complexity.
    \item With the enabled cooperation among all interfaces of the AP MLD, we propose a cooperative approach, where each interface employs the aggregate information across all interfaces upon gaining a TXOP under an EMLSR operation. With respect to the proposed cooperative approach, we describe its ISAC decision and STA MLD selection procedures, and we give an analysis of its computational complexity.
\end{itemize}

A preliminary conference version of this paper appears in \cite{tai2024d}.
For the initial conference paper, the proposed approach is less generalizable without systematic design principles.
In this paper, we have made major updates and have included more contents significantly.
First, we systematically present key design principles of ISAC decision and STA MLD selection.
We not only develop a new time-based strategy to determine an action between sensing and communications regarding ISAC decision but also provide a new theorem with an analysis of CRLB-based trilateration performance metric and develop a new candidate strategy regarding STA MLD selection for UL sensing.
Second, we propose novel non-cooperative and cooperative approaches according to our key design principles.
These two proposed approaches take different philosophies: The proposed non-cooperative approach allows each interface of the AP MLD to leverage its own information, while the proposed cooperative approach has each interface of the AP MLD leverage the aggregate information across all interfaces.

The remainder of this paper is organized as follows.
In Sec. \ref{sec:system}, we outline the system model and problem formulation.
In Sec. \ref{sec:design}, we present the key design principles.
We propose the non-cooperative and cooperative approaches in Secs. \ref{sec:Non-Coop} and \ref{sec:Coop}, respectively.
Simulation settings and results are covered in Sec. \ref{sec:simulation}.
Finally, Sec. \ref{sec:conclusion} concludes the paper.

\emph{Symbols:} Boldfaced capital and lowercase letters denote matrices and column vectors, respectively.
Given a vector $\mathbf{a}$, we use $\mbox{diag}(\mathbf{a})$ to denote the diagonal matrix containing $\mathbf{a}$ on its diagonal.
Given a matrix $\mathbf{A}$, we denote $\mbox{Tr}\{\mathbf{A}\}$, $\mathbf{A}^T$, and $\mathbf{A}^{-1}$ its trace, transpose, and inverse, respectively.
For any matrices $\mathbf{A}$ and $\mathbf{B}$, we use $\mathbf{A}\otimes\mathbf{B}$ to denote their Kronecker product.
We define $\mathbf{I}_p$ to be the $p\times p$ identity matrix and use $\mathbf{S}_{++}^p$ to denote the set of symmetric positive definite $p\times p$ matrices.
For any set $\mathcal{A}$, we use $[\mathcal{A}]^p$ to denote its $p$-subsets.
We denote the multivariate normal distribution with mean vector $\mathbf{\mu}$ and covariance matrix $\mathbf{\Phi}$ as $\mathcal{N}(\mathbf{\mu},\mathbf{\Phi})$.

The main notations used in this paper are summarized in Table \ref{tab_notations}.

\begin{table}[h]
    \centering
    \caption{Main Notations}
    \label{tab_notations}
    \begin{tabular}{c|c}
    \hline
    Notation     & Definition\\
    \hline\hline
    $(b^r_m,b^x_m)$ & \# bytes that have been received and to be transmitted DL\\
    \hline
    $C_{r_{m,l}}$ & CRLB of range estimate\\
    \hline
    $d_m$ ($\hat{d}_m$) & Distance between target (predicted position) and STA MLD\\
    \hline
    $\mathbf{F}$ & Transition matrix for target motion\\
    \hline
    $\mathbf{g}$ & Process noise\\
    \hline
    $g_s$ & Process noise intensity\\
    \hline
    $\mathbf{H}$ & Mapping matrix for target position\\
    \hline
    $\mathcal{I}^a_l$/$\mathcal{I}^c_l$/$\mathcal{I}_l$ & Set of available/candidate/selected STA MLD indices\\
    \hline
    $k$ & Maximum \# candidates\\
    \hline
    $\mathbf{K}$ & Kalman gain matrix\\
    \hline
    $L$ & \# interfaces in MLD\\
    \hline
    $M$ & \# STA MLDs\\
    \hline
    $\mathcal{M}$ & Set of STA MLD indices\\
    \hline
    $N$ & \# previous sensing TXOPs in current time window\\
    \hline
    $p_l$ & Maximum \# bytes to be transmitted DL\\
    \hline
    $t$/$t'$ & Time at which current TXOP/last sensing TXOP occurs\\
    \hline
    $t_E$ & End time of current time window\\
    \hline
    $t^*$ & Time-based criterion between sensing and communications\\
    \hline
    $T'$ & Elapsed time from last sensing TXOP\\
    \hline
    $T^{\#}$ & Remaining time in current time window\\
    \hline
    $\mbox{Tr}\{\hat{\mathbf{\Psi}}^{-1}_{\mathcal{I}_l,l}\}$ & Predicted CRLB of trilateration estimate \\
    \hline
    $\mathbf{v}_{\mathcal{I}_l}$ & Measurement noise\\
    \hline
    $w_m$ & Assigned weight\\
    \hline
    $\mathbf{x}$/$\mathbf{x}'$/$\hat{\mathbf{x}}$/$\tilde{\mathbf{x}}$/$\tilde{\mathbf{x}}'$ & Current/last/predicted/updated/last updated state of target\\
    \hline
    $(x,y)$/$(\hat{x},\hat{y})$ & Current/predicted position of target\\
    \hline
    $(\dot{x},\dot{y})$ & Current velocity of target\\
    \hline
    $(\bar{x}_m,\bar{y}_m)$ & Position of STA MLD\\
    \hline
    $\mathbf{z}$ & Trilateration measurement\\
    \hline
    $\alpha$ & Control variable between sensing and communications\\
    \hline
    $\beta$ & Binary indicator (1: sensing, 0: communications)\\
    \hline
    $\hat{\mathbf{\Delta}}$/$\tilde{\mathbf{\Delta}}$/$\tilde{\mathbf{\Delta}}'$ & Prediction/update/last update MSE matrix\\
    \hline
    $\eta$ & \# EHT-LTF repetitions in SR2SI NDP\\
    \hline
    $(\theta_s,\theta_c)$ & Sensing score and communications score\\
    \hline
    $(\xi^u_{m,l},\xi^d_{m,l})$ & UL and DL SNR\\
    \hline
    $\tau_{c,min}$/$\tau_{s,min}$ & Minimum time duration required for communications/sensing\\
    \hline
    $\tau_w$ & Duration of time window\\
    \hline
    $\psi_m$ & Average weighted utility per byte\\
    \hline
    $\omega_l$ & Signaling bandwidth\\
     \hline
    \end{tabular}
\end{table}

\section{System Model and Problem Formulation}
\label{sec:system}
In this section, we describe the system model and problem formulation of target tracking with ISAC using EMLSR in IEEE 802.11 WLANs.

\begin{figure}[ht]
\centering
\includegraphics[width=8.5cm]{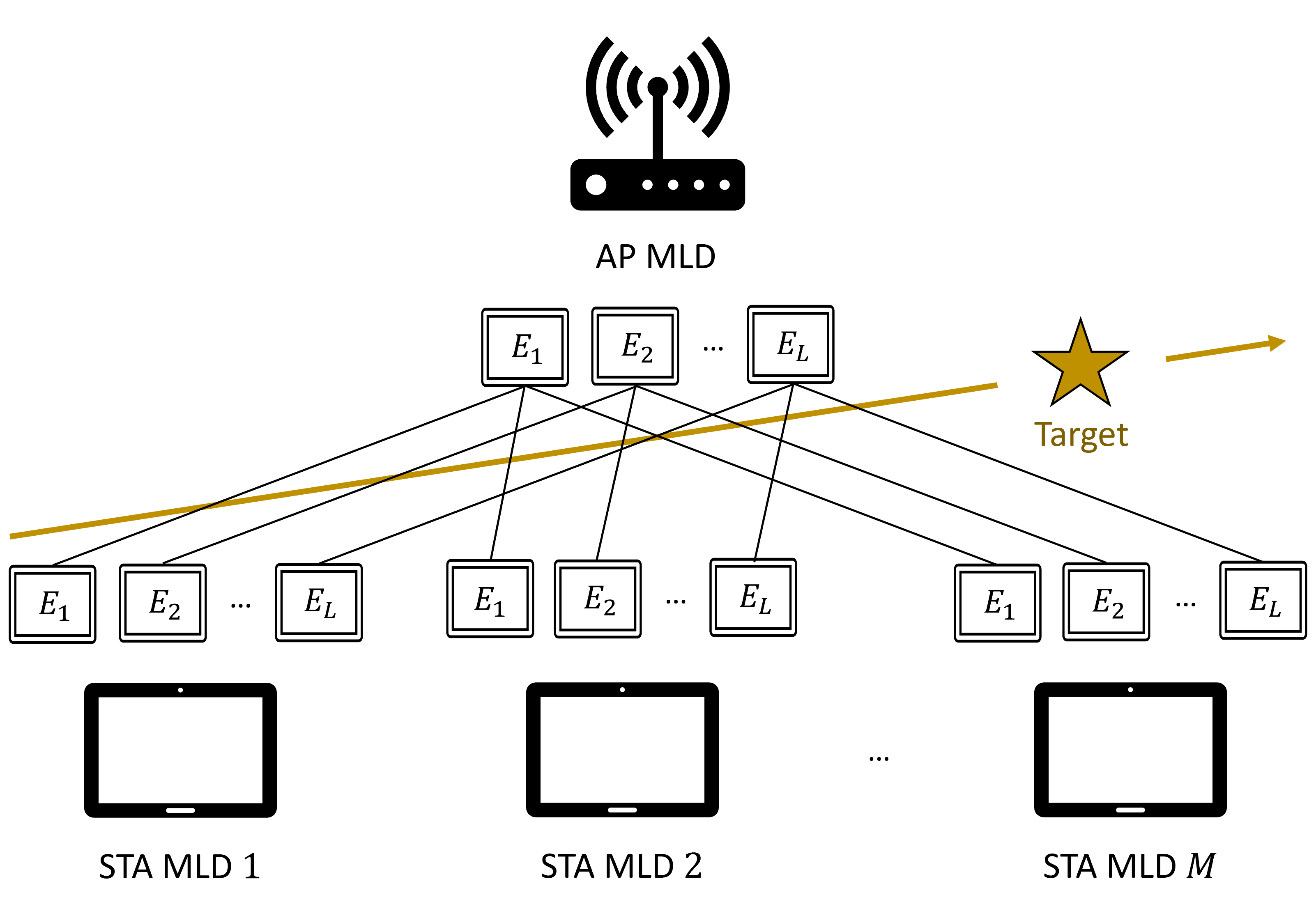}
\caption{An illustration of Wi-Fi network, where $E_l$ represents the $l$th interface and each solid line between interfaces represents a link}
\label{fig:network}
\end{figure}

Consider a Wi-Fi network with an AP MLD, $M$ STA MLDs, and a moving target (to be tracked) on a 2D area, where every MLD hosts $L$ interfaces.
Each STA MLD connects its $l$th interface to the $l$th interface of the AP MLD via its $l$th link over the $l$th channel at the $l$th frequency band, $l=1,2,\dots,L$.
Denote the set of STA MLD indices as $\mathcal{M}=\{1,2,\dots,M\}$.
An illustration of the Wi-Fi network is shown in Fig. \ref{fig:network}.
Furthermore, the Wi-Fi network features both EMLSR operation (supported by IEEE 802.11be) and ISAC functionality (supported by IEEE 802.11bf), with uplink (UL) sensing and downlink (DL) communications.

For EMLSR, define a time window as a period of time of duration $\tau_w$ accommodating multiple EMLSR operations, where each EMLSR operation takes place between the AP MLD and STA MLD(s) with two phases: link listening and frame exchange.
An EMLSR operation starts with the link listening phase.
Suppose at time $t$, the $l$th interface of the AP MLD gains a TXOP, and a set of STA MLDs of indices $\mathcal{I}^a_l\subseteq\mathcal{M}$ are listening to their respective $L$ links.
For the $m$th STA MLD, $m\in\mathcal{I}^a_l$, denote its position as $(\bar{x}_m,\bar{y}_m)$, the number of its bytes that have been received and to be transmitted DL as $b^r_m$ and $b^x_m$, respectively, and the UL and DL signal-to-noise ratio (SNR) of its $l$th link connected to the AP MLD as $\xi^u_{m,l}$ and $\xi^d_{m,l}$, respectively.
Then, the AP MLD selects some STA MLDs of indices $\mathcal{I}_l\subseteq\mathcal{I}^a_l$ to be involved, sending a multi-user request to send (MU-RTS) Trigger frame (TF) from its $l$th interface to the $l$th interface of each STA MLD belonging to $\mathcal{I}_l$.
Upon the reception of clear to send (CTS) frames responded from the STA MLDs, the AP MLD initiates the frame exchange phase.
After the frame exchange phase completes, the link listening phase resumes and a new EMLSR operation begins.
By the end of a time window, any ongoing EMLSR operation should finish.

Combined with ISAC, the AP MLD decides between sensing and communications under an EMLSR operation toward the TXOP gained by its $l$th interface at time $t$, given sufficient remaining time in the time window.
At time $t$, we express the current state of the target as
\begin{equation}
    \mathbf{x}=[x\,\dot{x}\,y\,\dot{y}]^T,
\end{equation}
where $(x,y)$ and $(\dot{x},\dot{y})$ are the current position and velocity of the target, respectively.
Define a binary variable $\beta$ of value 1 or 0 when the AP MLD decides to conduct sensing or communications, respectively.
For the current TXOP at time $t$, the AP MLD needs to generate a predicted state $\hat{\mathbf{x}}$ of the target and determine the value of $\beta\in\{0,1\}$.

\begin{figure}[h]%
\centering
\subfigure[UL sensing under EMLSR operation]{%
\label{fig:sensing}%
\includegraphics[width=8.5cm]{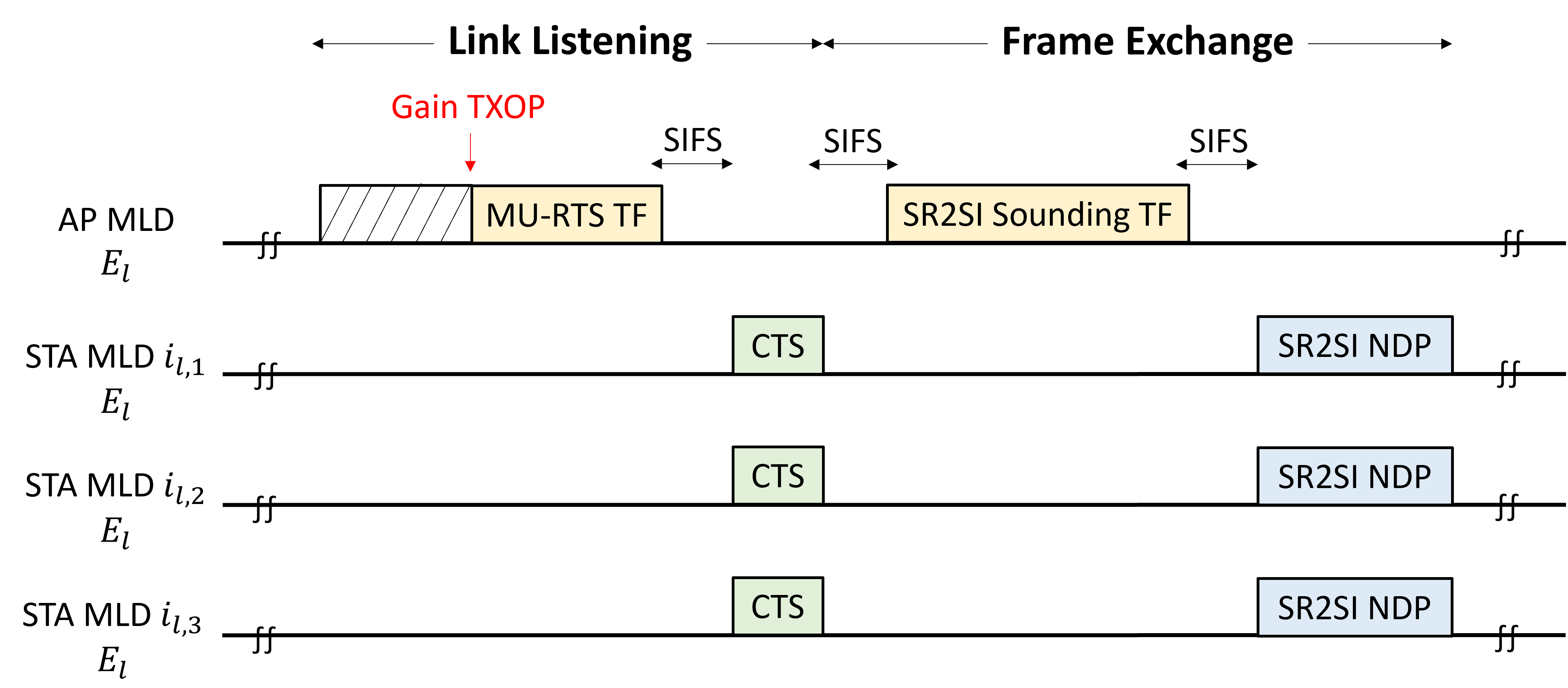}}%
\qquad
\subfigure[DL communications under EMLSR operation with $|\mathcal{I}_l|=2$]{%
\label{fig:comm}%
\includegraphics[width=8.5cm]{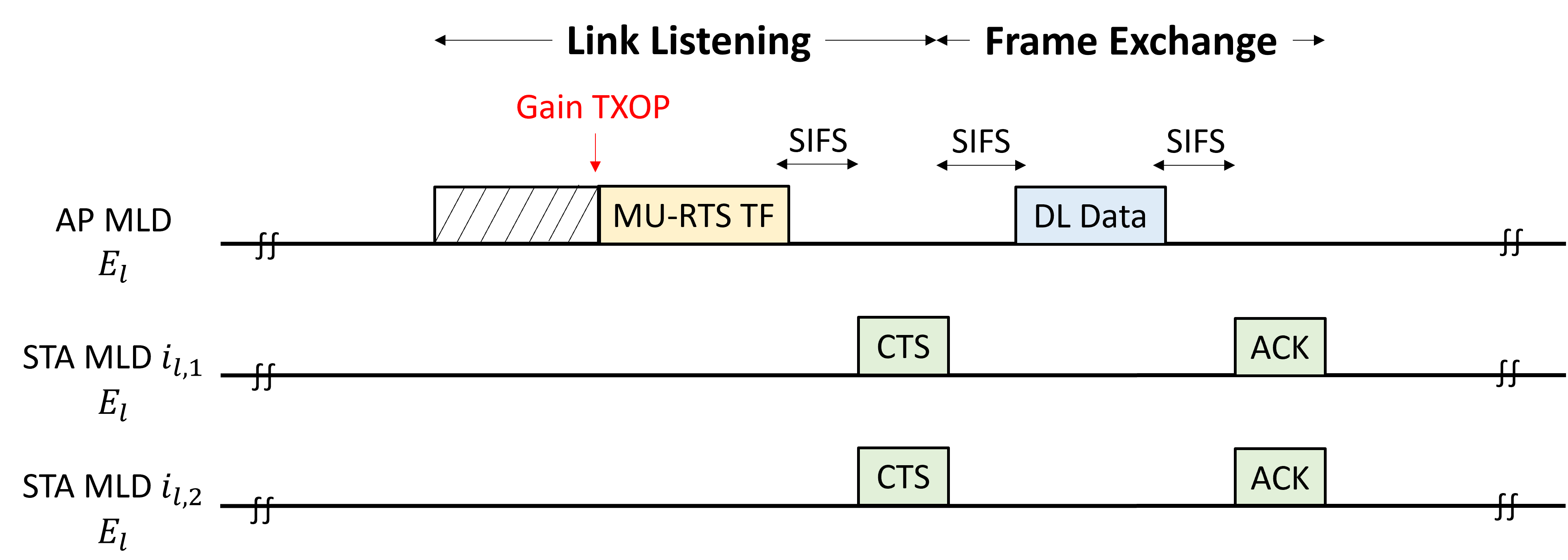}}%
\caption{An illustration of ISAC under EMLSR operation}
\label{fig:EMLSR}
\end{figure}

If the AP MLD decides to conduct sensing ($\beta=1$), then it will experience the current sensing TXOP with UL sensing to obtain a measurement for tracking the current state $\mathbf{x}$ of the target.
Considering the last sensing TXOP that occurred at time $t'$, the time duration between occurrence of the last and current sensing TXOPs is $T'=t-t'$.
Then, the target state transition between the last and current sensing TXOPs can be expressed with nearly constant velocity (CV) model \cite{barshalom2000} as
\begin{equation}
    \mathbf{x}=\mathbf{F}\mathbf{x}'+\mathbf{g},
\end{equation}
where $\mathbf{x}'$ is the last state of the target, $\mathbf{F}=\mathbf{I}_2\otimes\begin{bmatrix}
        1 & T' \\
        0 & 1
    \end{bmatrix}$, and $\mathbf{g}\sim\mathcal{N}(\mathbf{0},\mathbf{Q}_g)$ is the process noise with $\mathbf{Q}_g=g_s\mathbf{I}_2\otimes\begin{bmatrix}
    T'^3/3 & T'^2\\
    T'^2 & T'
\end{bmatrix}$ of process noise intensity $g_s$.
In UL sensing, the AP MLD (with its $l$th interface serving as SI) involves three STA MLDs (with their $l$th interface serving as SR) for three range estimates, where each range estimate results from the interaction between the AP MLD and an STA MLD, executing trilateration to obtain a measurement in terms of target position, as illustrated in Fig. \ref{fig:sensing}.
During the link listening phase, the AP MLD selects three STA MLDs of indices $\mathcal{I}_l=\{i_{l,1},i_{l,2},i_{l,3}\}\in[\mathcal{I}^a_l]^3$.
Next, during the frame exchange phase, the AP MLD sends an SR2SI Sounding TF to each STA MLD belonging to $\mathcal{I}_l$.
After receiving three SR2SI null data packets (NDPs) responded from the three STA MLDs, the AP MLD derives three range estimates for trilateration to obtain a measurement in terms of target position, expressed as
\begin{equation}
    \mathbf{z}=\mathbf{H}\mathbf{x}+\mathbf{v}_{\mathcal{I}_l},
\label{eq:z}
\end{equation}
where $\mathbf{H}=\begin{bmatrix}
        1 & 0 & 0 & 0\\
        0 & 0 & 1 & 0
    \end{bmatrix}$ and $\mathbf{v}_{\mathcal{I}_l}\sim\mathcal{N}(\mathbf{0},\mathbf{Q}_{v_{\mathcal{I}_l}})$ is the measurement noise with $\mathbf{Q}_{v_{\mathcal{I}_l}}=\mbox{diag}([\sigma_{\mathcal{I}_l}^2\,\sigma_{\mathcal{I}_l}^2]^T)$ of noise variance $\sigma_{\mathcal{I}_l}^2$ dependent on $\mathcal{I}_l$.

If the AP MLD decides to conduct communications ($\beta=0$), then it will experience the current communications TXOP with DL communications, as illustrated in Fig. \ref{fig:comm}.
During the link listening phase, the AP MLD selects some STA MLDs of indices $\mathcal{I}_l=\{i_{l,1},i_{l,2},\dots,i_{l,|\mathcal{I}_l|}\}$.
Next, during the frame exchange phase, the AP MLD sends DL data to each STA MLD belonging to $\mathcal{I}_l$, which responds with an acknowledgment (ACK) frame.

Accordingly, we formulate the following main problem of target tracking with ISAC using EMLSR in IEEE 802.11 WLANs:
For the TXOP gained by its $l$th interface under an EMLSR operation with $\mathcal{I}^a_l\subseteq\mathcal{M}$ at time $t$, given STA MLD position $\{(\bar{x}_m,\bar{y}_m)\}_{m\in\mathcal{I}^a_l}$, number of bytes that have been received and to be transmitted DL $\{(b^r_m,b^x_m)\}_{m\in\mathcal{I}^a_l}$, and UL and DL SNR of the $l$th link $\{(\xi^u_{m,l},\xi^d_{m,l})\}_{m\in\mathcal{I}^a_l}$ (along with history information from previous TXOPs), the AP MLD needs to make:
\begin{enumerate}
    \item \emph{ISAC decision}: Generate a predicted state $\hat{\mathbf{x}}$ of the target and determine the value of $\beta\in\{0,1\}$.
    \item \emph{STA MLD selection}: Determine the indices of STA MLDs to be selected $\mathcal{I}_l\subseteq\mathcal{I}^a_l$.
\end{enumerate}
Consequently, we focus on how the AP MLD should appropriately make an ISAC decision and an STA MLD selection for each TXOP in this paper.
The pivotal design principles of ISAC decision and STA MLD selection will be discussed in the next section.

\section{Design Principles of ISAC Decision and STA MLD Selection}
\label{sec:design}
In this section, we elaborate on our key design principles of ISAC decision and STA MLD selection for the AP MLD, based on the main problem of target tracking with ISAC using EMLSR in IEEE 802.11 WLANs formulated in Sec. \ref{sec:system}.
First, we deal with ISAC decision with two goals, tracking the state of the target and determining whether to conduct sensing or communications.
Next, we deal with STA MLD selection for UL sensing and for DL communications, respectively.

\subsection{ISAC Decision}
With respect to an ISAC decision, the AP MLD needs to track the target state and determines its action between sensing and communications.

To track the target state, we adopt the Kalman filter \cite{kim2018introduction} with two steps: prediction and update.
Regarding the target, the AP MLD generates a predicted state (accompanied by the prediction mean squared error (MSE) matrix) with the prediction step for each TXOP and an updated state (accompanied by the update MSE matrix) with the update step for each sensing TXOP.
Toward the current TXOP at time $t$, the AP MLD computes a predicted state of the target (with updated state $\tilde{\mathbf{x}}'$ and update MSE matrix $\tilde{\mathbf{\Delta}}'$ from the last sensing TXOP that occurred at time $t'$) as
\begin{equation}
    \hat{\mathbf{x}}=\mathbf{F}\tilde{\mathbf{x}}'=[\hat{x}\ \hat{\dot{x}}\ \hat{y}\ \hat{\dot{y}}]^T,
\label{eq:hat_x}
\end{equation}
where $(\hat{x},\hat{y})$ and $(\hat{\dot{x}},\hat{\dot{y}})$ are the predicted position and velocity of the target, respectively, accompanied by the prediction MSE matrix $\hat{\mathbf{\Delta}}=\mathbf{F}\tilde{\mathbf{\Delta}}'\mathbf{F}^T+\mathbf{Q}_g$.
If the AP MLD decides to conduct sensing ($\beta=1$) and obtains a measurement $\mathbf{z}$ with (\ref{eq:z}), then it will compute the updated state of the target as
\begin{equation}
    \tilde{\mathbf{x}}=\hat{\mathbf{x}}+\mathbf{K}(\mathbf{z}-\mathbf{H}\hat{\mathbf{x}}),
\label{eq:tilde_x}
\end{equation}
where $\mathbf{K}=\hat{\mathbf{\Delta}}\mathbf{H}^T(\mathbf{Q}_{v_{\mathcal{I}_l}}+\mathbf{H}\hat{\mathbf{\Delta}}\mathbf{H}^T)^{-1}$ is the Kalman gain matrix, accompanied by the update MSE matrix $\tilde{\mathbf{\Delta}}=(\mathbf{I}_4-\mathbf{K}\mathbf{H})\hat{\mathbf{\Delta}}$.
Therefore, the AP MLD generates a predicted state $\hat{\mathbf{x}}$ with (\ref{eq:hat_x}) for the current TXOP at time $t$ and an updated state $\tilde{\mathbf{x}}$ with (\ref{eq:tilde_x}) if sensing is conducted ($\beta=1$).

To determine an action between sensing and communications, we develop a time-based strategy.
For the current TXOP at time $t$, suppose the last sensing TXOP occurred at time $t'$ and the current time window will end at time $t_E$.
Then, we define the following two instances of time duration:
\begin{enumerate}
    \item \emph{Elapsed time from last sensing TXOP}: Difference between previous time $t'$ and current time $t$, expressed as $T'=t-t'$ (following the same definition in Sec. \ref{sec:system})
    \item \emph{Remaining time in current time window}: Difference between current time $t$ and end time $t_E$, expressed as $T^{\#}=t_E-t$
\end{enumerate}
The AP MLD requires a minimum time duration
\begin{equation}
\tau_{s,min}=3\tau_{SIFS}+2\tau_{TF}+\tau_{CTS}+\tau_{NDP}
\end{equation} 
to conduct UL sensing from Fig. \ref{fig:sensing} (same time consumption for each sensing TXOP) and a minimum time duration
\begin{equation}
\tau_{c,min}=3\tau_{SIFS}+\tau_{TF}+\tau_{CTS}+\tau_{NDP}+\tau_{ACK}
\end{equation}
to conduct DL communications from Fig. \ref{fig:comm} (varying time consumption for each communications TXOP depending on DL data), where $\tau_{SIFS}$, $\tau_{TF}$, $\tau_{CTS}$, $\tau_{NDP}$, and $\tau_{ACK}$ are the time duration of short interframe space (SIFS), TF transmission, CTS transmission, NDP transmission, and ACK transmission, respectively.
To ensure proper functioning, the AP MLD determines an action between sensing and communications (i.e., determine the value of $\beta\in{0,1}$) only when the set
\begin{equation}
    \mathcal{I}^a_l\neq\varnothing
    \label{eq:cond_nonempty}
\end{equation}
and the remaining time
\begin{equation}
    T^{\#}\geq\tau_{min}=\mbox{max}\{\tau_{s,min},\tau_{c,min}\}.
    \label{eq:cond_time}
\end{equation}
Define a control variable $\alpha\in(0,1)$ for a tradeoff between sensing and communications, where a large $\alpha$ value is favorable for sensing and a small $\alpha$ value is favorable for communications.
Note that in a single time window, an additional sensing measurement \emph{on average} introduces less correction for the target state (due to a decrease in average elapsed time) but incurs the same reduction in the communications throughput (due to the same time consumption of each sensing TXOP).
Therefore, we decrease the tendency to sensing as the number of previous sensing TXOPs in the current time window increases.
Since a larger value of $T'$ encourages sensing more (due to decreased fidelity of last measurement) and a larger value of $T^{\#}$ encourages communications more (due to increased capability of data transmission), we define the sensing score and the communications score as
\begin{equation}
    (\theta_s,\theta_c)=(\alpha^{N+1}T',(1-\alpha^{N+1})T^{\#}),
\end{equation}
where $N$ is the number of previous sensing TXOPs in the current time window.
For an action between sensing and communications, we obtain
\begin{equation}
    \beta=\mathbf{1}_{\{|\mathcal{I}_l^a|\geq 3\ \mbox{\small{\textbf{and}}}\ \theta_s>\theta_c\}}(|\mathcal{I}_l^a|,\theta_s,\theta_c),
    \label{eq:beta_theta}
\end{equation}
which is equal to 1 (sensing) if $|\mathcal{I}_l^a|\geq 3$ (as trilateration requires three STA MLDs) and $\theta_s>\theta_c$ or equal to 0 (communications) otherwise.
With some elementary mathematical manipulations, it can be shown that the condition of $\theta_s>\theta_c$ is equivalent to the condition of $t>t^*$, where
\begin{equation}
    t^*=\alpha^{N+1}t'+(1-\alpha^{N+1})t_E.
    \label{eq:t_star}
\end{equation}
Note that $t^*$ in (\ref{eq:t_star}) will be updated if the last TXOP is a sensing TXOP (due to updated $t'$).
Accordingly, (\ref{eq:beta_theta}) can be rewritten as
\begin{equation}
    \beta=\mathbf{1}_{\{|\mathcal{I}_l^a|\geq 3\ \mbox{\small{\textbf{and}}}\ t>t^*\}}(|\mathcal{I}_l^a|,t).
    \label{eq:beta_t}
\end{equation}
In (\ref{eq:beta_t}), we provide a clear time-based criterion (along with the trilateration requirement) for the AP MLD to determine an action between sensing and communications.

\subsection{STA MLD Selection for UL Sensing}
When the AP MLD decides to conduct sensing ($\beta=1$), it needs to select three STA MLDs of indices $\mathcal{I}_l=\{i_{l,1},i_{l,2},i_{l,3}\}\in[\mathcal{I}^a_l]^3$ for trilateration to obtain a measurement in UL sensing.

To quantify the trilateration performance of the AP MLD, we develop a trilateration performance metric based on CRLB \cite{nielsen2013cramer}, which indicates the minimum variance of an unbiased estimate.
Following the CRLB analysis in \cite{richards2022}, we obtain the CRLB of range estimate and trilateration estimate for UL sensing (from Fig. \ref{fig:sensing}).
To begin with, the CRLB of range estimate between the AP MLD and the $m$th STA MLD (with their $l$th interface) can be obtained as
\begin{equation}
C_{r_{m,l}}=\mu/(\omega_l^2\xi_{m,l}^u),
\label{eq:CRLB_range}
\end{equation}
where $\mu=3c^2/(8\pi^2\eta)$ with $c$ being the speed of light and $\eta$ being the number of EHT-long training field (EHT-LTF) repetitions in each SR2SI NDP, and $\omega_l$ is the signaling bandwidth in the $l$th channel.
With an extension of (\ref{eq:CRLB_range}), the CRLB of trilateration estimate between the AP MLD and the three STA MLDs of indices $\mathcal{I}_l=\{i_{l,1},i_{l,2},i_{l,3}\}$ (with their $l$th interface) can be obtained as
\begin{equation}
    C_{t_{\mathcal{I}_l,l}}=\mbox{Tr}\{\mathbf{\Psi}_{\mathcal{I}_l,l}^{-1}\}=\mbox{Tr}\{(\mathbf{\Gamma}_{\mathcal{I}_l,l}\mathbf{D}_{\mathcal{I}_l,l}\mathbf{\Gamma}_{\mathcal{I}_l,l}^T)^{-1}\},
    \label{eq:CRLB_tri}
\end{equation}
where $\mathbf{D}_{\mathcal{I}_l,l}=\mbox{diag}([C_{r_{i_{l,1},l}}^{-1}\,C_{r_{i_{l,2},l}}^{-1}\,C_{r_{i_{l,3},l}}^{-1}]^T)$ and $\mathbf{\Gamma}_{\mathcal{I}_l,l}=\begin{bmatrix}
    (x-\bar{x}_{i_{l,1}})/d_{i_{l.1}} & (x-\bar{x}_{i_{l,2}})/d_{i_{l,2}} & (x-\bar{x}_{i_{l,3}})/d_{i_{l,3}} \\
    (y-\bar{y}_{i_{l,1}})/d_{i_{l,1}} & (y-\bar{y}_{i_{l,2}})/d_{i_{l,2}} & (y-\bar{y}_{i_{l,3}})/d_{i_{l,3}}
\end{bmatrix}$
with $(x,y)$ being the current target position and $d_{i_{l,j}}=\sqrt{(x-\bar{x}_{i_{l,j}})^2+(y-\bar{y}_{i_{l,j}})^2}$ being the distance between the target and the $i_{l,j}$th STA MLD, $j=1,2,3$.
Since the current target position $(x,y)$ is unknown to the AP MLD, we replace $(x,y)$ in (\ref{eq:CRLB_tri}) with the predicted target position in (\ref{eq:hat_x}) and obtain the predicted CRLB of trilateration estimate as
\begin{equation}
    \hat{C}_{t_{\mathcal{I}_l,l}}=\mbox{Tr}\{\hat{\mathbf{\Psi}}_{\mathcal{I}_l,l}^{-1}\}=\mbox{Tr}\{(\hat{\mathbf{\Gamma}}_{\mathcal{I}_l,l}\mathbf{D}_{\mathcal{I}_l,l}\hat{\mathbf{\Gamma}}_{\mathcal{I}_l,l}^T)^{-1}\},
    \label{eq:hat_CRLB_tri}
\end{equation}
where $\hat{\mathbf{\Gamma}}_{\mathcal{I}_l,l}=\mathbf{\Gamma}_{\mathcal{I}_l,l}|_{(x,y)\leftarrow(\hat{x},\hat{y})}$ with $\hat{d}_{i_{l,j}}=d_{i_{l,j}}|_{(x,y)\leftarrow(\hat{x},\hat{y})}$.
Note that as $\hat{\mathbf{\Gamma}}_{\mathcal{I}_l,l}$ is a full rank matrix and $\mathbf{D}_{\mathcal{I}_l,l}$ is a diagonal matrix with all diagonal entries being positive, $\hat{\mathbf{\Psi}}_{\mathcal{I}_l,l}$ is a symmetric positive definite matrix.
Accordingly, we employ the predicted CRLB of trilateration estimate $\hat{C}_{t_{\mathcal{I}_l,l}}$ in (\ref{eq:hat_CRLB_tri}) as our trilateration performance metric for the AP MLD.

Therefore, the AP MLD achieves the best trilateration performance in UL sensing with a selection of three STA MLDs of indices $\mathcal{I}_l\in[\mathcal{I}_l^a]^3$ leading to minimum predicted CRLB of trilateration estimate $\hat{C}_{t_{\mathcal{I}_l,l}}=\mbox{Tr}\{\hat{\mathbf{\Psi}}_{\mathcal{I}_l,l}^{-1}\}$ in (\ref{eq:hat_CRLB_tri}), which is equivalent to solving the optimization problem (\ref{DCO}) below:
\begin{subequations}
\label{DCO}
\begin{align}
\underset{\hat{\mathbf{\Psi}}\in\mathbf{S}_{++}^2}{\mbox{min}}\,\ \quad  & \mbox{Tr}\{\hat{\mathbf{\Psi}}^{-1}\}\label{eq:objective}\\ 
\mbox{subject to}\quad & \hat{\mathbf{\Psi}}\in\{\hat{\mathbf{\Psi}}_{\mathcal{I}_l,l}:\mathcal{I}_l\in[\mathcal{I}_l^a]^3\}\label{eq:constraint}
\end{align}
\end{subequations}
The objective function (\ref{eq:objective}) is convex in $\hat{\mathbf{\Psi}}$ (since the trace function of the inverse of a symmetric positive definite matrix is convex), and the constraint (\ref{eq:constraint}) specifies the set of choices for $\hat{\mathbf{\Psi}}$.
As a result, the optimization problem (\ref{DCO}) is a discrete convex optimization problem, which can be solved with existing techniques (e.g., \cite{valls2019,Murota2009}).

However, when $|\mathcal{I}^a_l|$ is large, directly solving the discrete convex optimization problem (\ref{DCO}) involves high computational complexity due to the broad search space.
As an alternative, we further investigate the discrete convex optimization problem (\ref{DCO}) and devise a feasible strategy.

For a deeper understanding of the discrete convex optimization problem (\ref{DCO}), we theoretically analyze the objective function (\ref{eq:objective}).
The theorem below provides a lower bound of the objective function (\ref{eq:objective}) with the predicted CRLB of trilateration estimate under the constraint (\ref{eq:constraint}) with the specified set of choices.

\textit{Theorem 1 (Lower bound of predicted CRLB of trilateration estimate):} Given $\hat{\mathbf{\Psi}}$ under the constraint (\ref{eq:constraint}), the predicted CRLB of trilateration estimate $\mbox{Tr}\{\hat{\mathbf{\Psi}}^{-1}\}$ in the objective function (\ref{eq:objective}) satisfies
\begin{equation}
    \mbox{Tr}\{\hat{\mathbf{\Psi}}^{-1}\}\geq 4\mu/(\omega_l^2\xi_l^*),
\end{equation}
where $\xi_l^*=\underset{\mathcal{I}_l=\{i_{l,1},i_{l,2},i_{l,3}\}\in[\mathcal{I}^a_l]^3}{\mbox{max}}\sum_{j=1}^3 \xi^u_{i_{l,j},l}$.
\begin{proof}
See Appendix \ref{pf:thm1}.
\end{proof}
In Theorem 1, the lower bound of $\mbox{Tr}\{\hat{\mathbf{\Psi}}^{-1}\}$ in the objective function (\ref{eq:objective}) depends on the term $\xi_l^*$, which is the maximum sum of UL SNRs from the three STA MLDs of indices $\mathcal{I}_l\in[\mathcal{I}^a_l]^3$ under the constraint (\ref{eq:constraint}).
This implies that the three STA MLDs with high UL SNRs have much potential to result in good trilateration performance.

Despite the lower bound in Theorem 1, it is not guaranteed that the three STA MLDs with highest UL SNRs \emph{always} result in the best trilateration performance.
Therefore, we develop a $k$-candidates strategy, where we nominate up to $k$ candidate STA MLDs with highest UL SNRs and select three STA MLDs out of them with the best possible trilateration performance.
Given $\mathcal{I}^a_l$, the index set of candidate STA MLDs can be obtained as
{\fontsize{9}{6}
\begin{equation}
\mathcal{I}^c_l = \left\{\begin{array}{ll}
\underset{\mathcal{I}_l=\{i_{l,1},i_{l,2},\dots,i_{l,k}\}\in[\mathcal{I}^a_l]^k}{\mbox{argmax}}\sum_{j=1}^k\xi^u_{i_{l,j},l}, & |\mathcal{I}^a_l|>k\\
\mathcal{I}^a_l, & |\mathcal{I}^a_l|\leq k
\end{array}\right..
\label{eq:candidate}
\end{equation}
}According to (\ref{eq:candidate}), the $k$ candidates with highest UL SNRs are nominated from all STA MLDs belonging to $\mathcal{I}^a_l$ if $|\mathcal{I}^a_l|>k$, while all STA MLDs belonging to $\mathcal{I}^a_l$ are nominated as candidates if $|\mathcal{I}^a_l|\leq k$.
Note that the value of $k$ should be carefully chosen.
A too large $k$ value incurs an explosive increase in computational complexity, while a too small $k$ value leads to a narrow search space.
With the $k$-candidates strategy (on a proper $k$ value), we enable the AP MLD to select the three STA MLDs for UL sensing in a practical manner.

\subsection{STA MLD Selection for DL Communications}
When the AP MLD decides to conduct communications ($\beta=0$), it needs to select some STA MLDs of indices $\mathcal{I}_l\subseteq\mathcal{I}_l^a$ for data transmission in DL communications.

With both throughput and fairness taken into consideration, the AP MLD aspires to achieve weighted proportional fairness \cite{shi2014}, which is equivalent to solving the optimization problem (\ref{PF}) below:
\begin{subequations}
\label{PF}
\begin{align}
\underset{\mathcal{I}_l\subseteq\mathcal{I}_l^a}{\mbox{max}}\qquad & \sum_{m\in\mathcal{I}_l}w_m\,\mbox{log}(b^x_m)\label{eq:objective2}\\
\mbox{subject to} \quad &  \sum_{m\in\mathcal{I}_l}b^x_m\leq p_l\label{eq:constraint2}
\end{align}
\end{subequations}
The objective function (\ref{eq:objective2}) is the sum of weighted utility, where $w_m$ and $\mbox{log}(b^x_m)$ are the assigned weight and the anticipated utility (dependent on the number of bytes to be transmitted DL), respectively, of the $m$th STA MLD, and the constraint (\ref{eq:constraint2}) specifies the upper bound $p_l$ of number of bytes to be transmitted DL over the $l$th channel.
Consider the bytes to be transmitted DL of each STA MLD as an item of finite size, the anticipated utility as the value of an item, and the channel as a knapsack of limited volume.
Then, it can be inferred that the optimization problem (\ref{PF}) is an NP-hard knapsack problem \cite{PISINGER20052271}.

Generally, an NP-hard knapsack problem like (\ref{PF}) is intractable.
Therefore, we are motivated to develop a feasible heuristic strategy.
For the STA MLDs belonging to $\mathcal{I}^a_l$, we normalize their number of bytes that have been received DL $\{b^r_m\}_{m\in\mathcal{I}^a_l}$ into z-score $\{z_{s_m}\}_{m\in\mathcal{I}^a_l}$.
Based on its z-score $z_{s_m}$, the $m$th STA MLD is assigned a weight, expressed as
\begin{equation}
    w_m=\mbox{exp}(-z_{s_m}).
    \label{eq:weight}
\end{equation}
Namely, if an STA MLD has fewer bytes received DL (with a lower z-score), then it is assigned a larger weight.
With the aim of maximizing the sum of weighted utility in the objective function (\ref{eq:objective2}), an STA MLD with a larger average weighted utility per byte is given a higher priority in a greedy manner.
Specifically, the average weighted utility per byte of the $m$th STA MLD is computed as
\begin{equation}
    \psi_m=w_m\mbox{log}(b^x_m)/b^x_m.
    \label{eq:avg_weighted_utility}
\end{equation}
Then, we sort the average weighted utility per byte of the STA MLDs in descending order as $\psi_{j_1}\geq\psi_{j_2}\geq\dots\geq\psi_{j_{|\mathcal{I}^a_l|}}$ with $j_1,j_2,\dots,j_{|\mathcal{I}^a_l|}\in\mathcal{I}^a_l$ being the order of indices of STA MLDs to be addressed.
Following the order, the AP MLD addresses one STA MLD in an iteration (start from the $j_1$th STA MLD, then the $j_2$th STA MLD, and so on), selecting the STA MLD and accommodating its number of bytes to be transmitted DL, until the upper bound of number of bytes to be transmitted DL over the channel is reached.
With the feasible heuristic strategy, we provide a realistic scheme for the AP MLD to select STA MLDs for DL communications.

\begin{figure}[h]%
\centering
\subfigure[Non-cooperative approach]{%
\label{fig:illustration_NC}%
\includegraphics[width=8.5cm]{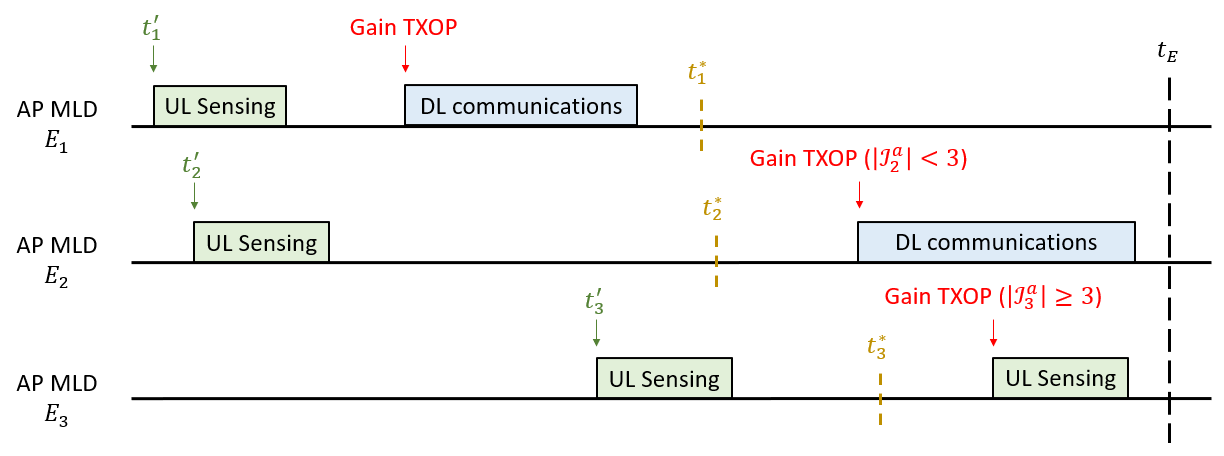}}%
\qquad
\subfigure[Cooperative approach]{%
\label{fig:illustration_C}%
\includegraphics[width=8.5cm]{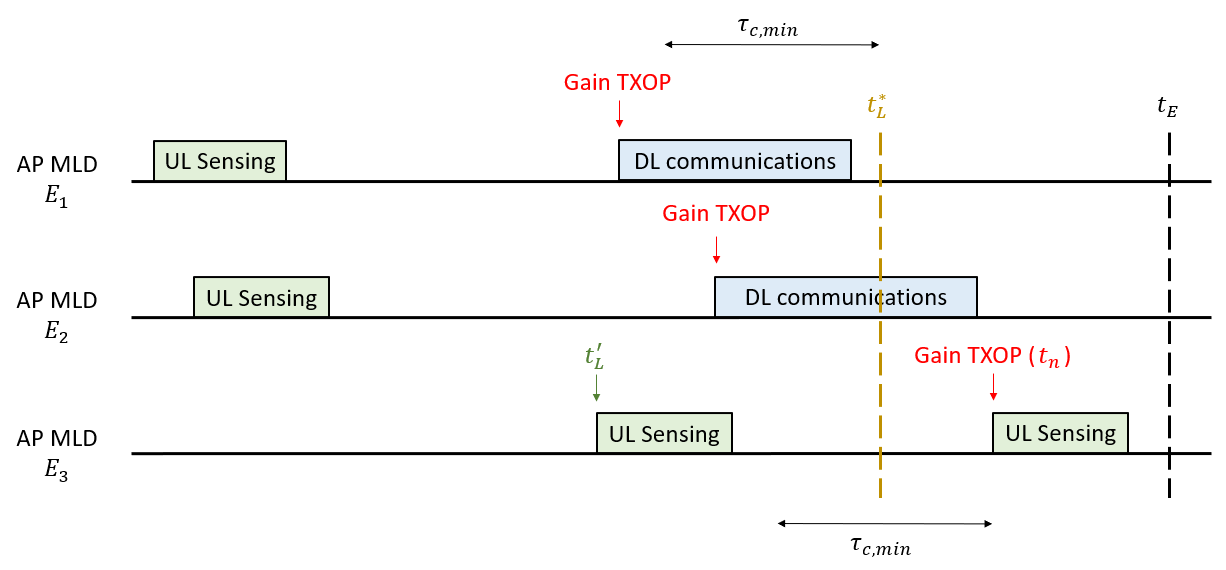}}%
\caption{An illustration of proposed non-cooperative and cooperative approaches with $N_1=N_2=N_3=N_L=0$ and $\alpha=0.5$}
\label{fig:approach}
\end{figure}

\section{Non-Cooperative Approach}
\label{sec:Non-Coop}
According to our key design principles elaborated in Sec. \ref{sec:design}, we would like to develop corresponding approaches toward ISAC decision and STA MLD selection for the AP MLD.
In this section, we propose a non-cooperative approach, where each interface of the AP MLD works with its own information (without the need of information from other interfaces) upon gaining a TXOP under an EMLSR operation.
For the proposed non-cooperative approach, we describe its procedures of ISAC decision and STA MLD selection in detail, and we analyze its computational complexity.
An illustration of the proposed non-cooperative approach is shown in Fig. \ref{fig:illustration_NC}.

\subsection{ISAC Decision}
\label{subsec:NC_ISAC_decision}
The proposed non-cooperative approach begins with the procedure of ISAC decision, which generates a predicted state of the target and determines an action between sensing and communications, as illustrated in Algorithm \ref{algo:NC_ISAC}.

\begin{algorithm}
\SetAlgoLined

\textbf{Input:} $\tilde{\mathbf{x}}_l',\mathcal{I}_l^a,t,t_l',t_E,N_l,\alpha$

\textbf{Initialization:} $T'=t-t_l',T^{\#}=t_E-t$

$\hat{\mathbf{x}}_l=\mathbf{F}\tilde{\mathbf{x}}_l'$

\textbf{if} $\mathcal{I}_l^a\neq\varnothing$ \textbf{and} $T^{\#}\geq\tau_{min}$

\quad $t_l^*=\alpha^{N_l+1}t_l'+(1-\alpha^{N_l+1})t_E$

\quad $\beta=\mathbf{1}_{\{|\mathcal{I}_l^a|\geq 3\ \mbox{\textbf{\small{and}}}\ t>t_l^*\}}(|\mathcal{I}_l^a|,t)$

\textbf{end if}

\textbf{Output:} $\hat{\mathbf{x}}_l,\beta$

\caption{Non-Cooperative ISAC Decision}
\label{algo:NC_ISAC}
\end{algorithm}

Associated with the $l$th interface, the last sensing TXOP occurred at time $t_l'$ with the updated state $\tilde{\mathbf{x}}_l'$ generated.
We initialize the elapsed time from last sensing TXOP $T'=t-t_l'$ and the remaining time in current time window $T^{\#}=t_E-t$.
When the $l$th interface gains a TXOP with $\mathcal{I}^a_l$ at time $t$, its own information is considered.
According to (\ref{eq:hat_x}), we compute a predicted state of the target as
\begin{equation}
    \hat{\mathbf{x}}_l=\mathbf{F}\tilde{\mathbf{x}}_l'.
    \label{eq:NC_predicted_state}
\end{equation}

If both conditions (\ref{eq:cond_nonempty}) and (\ref{eq:cond_time}) are met, we proceed to determine an action between sensing and communications.
With the number of previous sensing TXOPs in the current time window $N_l$ associated with the $l$th interface taken into account, we follow (\ref{eq:t_star}) and compute
\begin{equation}
    t_l^*=\alpha^{N_l+1}t_l'+(1-\alpha^{N_l+1})t_E.
    \label{eq:NC_tl_star}
\end{equation}
Substituting $t_l^*$ into $t^*$ in (\ref{eq:beta_t}), we obtain
\begin{equation}
    \beta=\mathbf{1}_{\{|\mathcal{I}_l^a|\geq 3\ \mbox{\textbf{\small{and}}}\ t>t_l^*\}}(|\mathcal{I}_l^a|,t),
    \label{eq:NC_beta}
\end{equation}
which is equal to 1 (sensing) if $|\mathcal{I}^a_l|\geq 3$ and $t>t_l^*$ or equal to 0 (communications) otherwise.

\subsection{STA MLD Selection}
\label{subsec:NC_STA_MLD_selection}
Depending on the value of $\beta$ in (\ref{eq:NC_beta}), the proposed non-cooperative approach executes the procedure of STA MLD selection for UL sensing ($\beta=1$) or for DL communications ($\beta=0$), as illustrated in Algorithm \ref{algo:NC_SMS}.

\begin{algorithm}
\SetAlgoLined

\textbf{Input:} $\beta$

\textbf{if} $\beta=1$

\quad \textbf{Input:} $\hat{\mathbf{x}}_l,\{\xi^u_{m,l}\}_{m\in\mathcal{I}^a_l},\{(\bar{x}_m,\bar{y}_m)\}_{m\in\mathcal{I}^a_l},k$

\quad \textbf{if} $|\mathcal{I}^a_l|>k$

{\fontsize{8}{6}
\qquad \textbf{Sort:} $\xi^u_{j_1,l}\geq\xi^u_{j_2,l}\geq\dots\geq\xi^u_{j_{|\mathcal{I}^a_l|},l},j_1,j_2,\dots,j_{|\mathcal{I}^a_l|}\in\mathcal{I}^a_l$
}

\qquad $\mathcal{I}^c_l=\{j_1,j_2,\dots,j_k\}$

\quad \textbf{else if} $|\mathcal{I}^a_l|\leq k$

\qquad $\mathcal{I}^c_l=\mathcal{I}^a_l$

\quad \textbf{end if}

\quad $\mathcal{I}_l=\underset{\mathcal{I}_l'\in[\mathcal{I}^c_l]^3}{\mbox{argmin}}\ \mbox{Tr}\{\hat{\mathbf{\Psi}}_{\mathcal{I}_l',l}^{-1}\}$

\quad \textbf{Trilateration measurement:} $\mathbf{z}$

\quad $\tilde{\mathbf{x}}_l=\hat{\mathbf{x}}_l+\mathbf{K}(\mathbf{z}-\mathbf{H}\hat{\mathbf{x}}_l)$

\textbf{Output:} $\mathcal{I}_l,\tilde{\mathbf{x}}_l$

\textbf{else if} $\beta=0$

\quad \textbf{Input:} $\{(b_m^r,b_m^x)\}_{m\in\mathcal{I}_l^a},p^E_l$

\quad \textbf{Initialization:} $\mathcal{I}_l=\varnothing,p'_l=p^E_l$

\quad \textbf{z-score normalization:} $\{b_m^r\}_{m\in\mathcal{I}_l^a}\rightarrow\{z_{s_m}\}_{m\in\mathcal{I}_l^a}$

\quad\textbf{for} $m\in\mathcal{I}_l^a$

\qquad $w_m=\mbox{exp}(-z_{s_m})$; $\psi_m=w_m\,\mbox{log}(b^x_m)/b^x_m$

\quad\textbf{end for}

{\fontsize{9}{6}
\quad\textbf{Sort:} $\psi_{j_1'}\geq \psi_{j_2'}\geq\dots\geq \psi_{j_{|\mathcal{I}_l^a|}'},j_1',j_2',\dots,j_{|\mathcal{I}_l^a|}'\in\mathcal{I}_l^a$
}

\quad\textbf{for} $q=1:|\mathcal{I}_l^a|$

\qquad $\mathcal{I}_l\leftarrow\mathcal{I}_l\cup j_q'$; $p'_l\leftarrow p'_l-b_{j_q'}^x$

\qquad\textbf{if} $p'_l<0$

\quad\qquad \textbf{break}

\qquad\textbf{end if}

\quad\textbf{end for}

\quad\textbf{Output:} $\mathcal{I}_l$

\textbf{end if}

\caption{Non-Cooperative STA MLD Selection}
\label{algo:NC_SMS}
\end{algorithm}

If an action of sensing is determined ($\beta=1$), then we are required to select three STA MLDs of indices $\mathcal{I}_l\in[\mathcal{I}^a_l]^3$ for trilateration in UL sensing.

Following (\ref{eq:candidate}) based on the developed $k$-candidates strategy, we obtain the index set of candidate STA MLDs $\mathcal{I}^c_l\subseteq\mathcal{I}^a_l$ according to the relationship between $|\mathcal{I}^a_l|$ and $k$.
If $|\mathcal{I}^a_l|>k$, then we sort the UL SNR of all STA MLDs belonging to $\mathcal{I}^a_l$ in descending order and nominate $k$ candidates with highest UL SNRs as $\mathcal{I}^c_l$.
If $|\mathcal{I}^a_l|\leq k$, then all STA MLDs belonging to $\mathcal{I}^a_l$ are nominated as candidates, i.e., $\mathcal{I}^c_l=\mathcal{I}^a_l$.
From the candidates in $\mathcal{I}^c_l$, we then select three STA MLDs with minimum predicted CRLB of trilateration estimate, whose indices can be expressed as
\begin{equation}
    \mathcal{I}_l=\underset{\mathcal{I}_l'\in[\mathcal{I}^c_l]^3}{\mbox{argmin}}\ \mbox{Tr}\{\hat{\mathbf{\Psi}}_{\mathcal{I}_l',l}^{-1}\}.
    \label{eq:sensing_STA_MLD}
\end{equation}
Employing the predicted state $\hat{\mathbf{x}}_l$ in (\ref{eq:NC_predicted_state}) and the trilateration measurement $\mathbf{z}$ obtained with the three STA MLDs of indices $\mathcal{I}_l$ in (\ref{eq:sensing_STA_MLD}) associated with the $l$th interface, we compute an updated state of the target, according to (\ref{eq:tilde_x}), as
\begin{equation}
    \tilde{\mathbf{x}}_l=\hat{\mathbf{x}}_l+\mathbf{K}(\mathbf{z}-\mathbf{H}\hat{\mathbf{x}}_l).
    \label{eq:NC_updated_state}
\end{equation}

If an action of communications is determined ($\beta=0$), then we are required to select some STA MLDs of indices $\mathcal{I}_l\subseteq\mathcal{I}^a_l$ for data transmission in DL communications.

Denote the remaining number of bytes that can be transmitted DL over the $l$th channel as $p_l'$.
We initialize
\begin{equation}
    \mathcal{I}_l=\varnothing
    \label{init:set}
\end{equation}
along with
\begin{equation}
    p_l'=p^E_l,
    \label{init:p_E}
\end{equation}
where $p^E_l$ is the maximum number of bytes to be transmitted DL over the $l$th channel by the end time of current time window $t_E$.
Then, we normalize the number of bytes that have been received DL $\{b^r_m\}_{m\in\mathcal{I}^a_l}$ into z-score $\{z_{s_m}\}_{m\in\mathcal{I}^a_l}$.
For the $m$th STA MLD, we assign its weight $w_m$, which is larger under a lower z-score $z_{s_m}$, with (\ref{eq:weight}) and compute its average weighted utility per byte $\psi_m$, which reflects its priority to be addressed, with (\ref{eq:avg_weighted_utility}).
By sorting the average weighted utility per byte of the STA MLDs in descending order as $\psi_{j_1'}\geq\psi_{j_2'}\geq\dots\geq\psi_{j_{|\mathcal{I}^a_l|}'}$, we obtain $j_1',j_2',\dots,j_{|\mathcal{I}^a_l|}'\in\mathcal{I}^a_l$ as the order of indices of STA MLDs to be addressed.
In the $q$th iteration, the AP MLD addresses the $j_q'$th STA MLD by adding the index $j_q'$ to $\mathcal{I}_l$ and subtracting the number of bytes to be transmitted DL $b^x_{j_q'}$ from $p_l'$.
Once $p_l'$ goes below zero, the AP MLD ends the STA MLD selection and obtains $\mathcal{I}_l$.

\subsection{Computational Complexity}
With its procedures of ISAC decision (Algorithm \ref{algo:NC_ISAC}) and STA MLD selection (Algorithm \ref{algo:NC_SMS}) as described in Secs. \ref{subsec:NC_ISAC_decision} and \ref{subsec:NC_STA_MLD_selection}, respectively, the proposed non-cooperative approach allows each interface of the AP MLD to work with its own information (not requiring information from other interfaces).
Below, we analyze the computational complexity of the proposed non-cooperative approach in terms of the number of multiplications/divisions involved.

First, we investigate the ISAC decision procedure (Algorithm \ref{algo:NC_ISAC}).
The computation of predicted state in (\ref{eq:NC_predicted_state}) involves two multiplications, since only two entries of $\mathbf{F}$ are non-binary (with value $T'>0$).
By maintaining (i.e., storing and keeping track of) the value of $N_l$ and $\alpha^{N_l+1}$ throughout the TXOPs associated with the $l$th interface, the update of $\alpha^{N_l+1}$ involves at most one multiplication with $\alpha$. 
Accordingly, the computation of $t_l^*$ in (\ref{eq:NC_tl_star}) involves at most three multiplications.
Overall, the computational complexity of the ISAC decision procedure (Algorithm \ref{algo:NC_ISAC}) is $\mathcal{O}(1)$.

Next, we investigate the STA MLD selection procedure (Algorithm \ref{algo:NC_SMS}).
We start with the case of UL sensing ($\beta=1$).
Given any $\mathcal{I}_l=\{i_{l,1},i_{l,2},i_{l,3}\}\in[\mathcal{I}^a_l]^3$, we analyze the computational complexity of deriving predicted CRLB of trilateration estimate in (\ref{eq:hat_CRLB_tri}) as follows.
Note that the computation of distance $\hat{d}_{i_{l,j}}$ between predicted position of target in (\ref{eq:NC_predicted_state}) and position of the $i_{l,j}$th STA MLD involves $\mathcal{O}(1)$ multiplications, $j=1,2,3$. 
Then, the computation of  $\hat{\mathbf{\Gamma}}_{\mathcal{I}_l,l}$ in (\ref{eq:Gamma_symbol}) involves six divisions (with each entry involving one division), and the computation of $\mathbf{D}_{\mathcal{I}_l,l}$ in (\ref{eq:D_symbol}) involves three multiplications (with each diagonal entry involving one multiplication by maintaining the value of $\omega_l^2/\mu$).
Referring to (\ref{eq:hat_Psi_symbol}), the computation of $\hat{\mathbf{\Psi}}_{\mathcal{I}_l,l}$ involves fifteen multiplications (with each $j$ index involving five multiplications resulting from $\gamma_{1,j}\rho_j$ and $\gamma_{2,j}\rho_j$ along with $\gamma_{1,j}^2\rho_j$, $\gamma_{2,j}^2\rho_j$, and $\gamma_{1,j}\gamma_{2,j}\rho_j$).
Then, the computation of inverse $\hat{\mathbf{\Psi}}_{\mathcal{I}_l,l}^{-1}$ involves two multiplications and one division.
Therefore, the computational complexity of deriving predicted CRLB of trilateration estimate in (\ref{eq:hat_CRLB_tri}) is $\mathcal{O}(1)$.
Accordingly, the computational complexity of selecting three STA MLDs (from up to $k$ candidates) with minimum predicted CRLB of trilateration estimate in (\ref{eq:sensing_STA_MLD}) is $\mathcal{O}(k^3)$.
The computation of updated state in (\ref{eq:NC_updated_state}) involves eight multiplications.
Overall, the computational complexity of the STA MLD selection procedure for UL sensing ($\beta=1$) is $\mathcal{O}(k^3)$. 
We move forward with the case of DL communications ($\beta=0$).
Note that the z-score normalization involves $\mathcal{O}(M)$ multiplications/divisions.
Then, the computation of weight in (\ref{eq:weight}) and average weighted utility per byte in (\ref{eq:avg_weighted_utility}) for the $\mathcal{O}(M)$ STA MLDs belonging to $\mathcal{I}^a_l$ involves $\mathcal{O}(M)$ multiplications/divisions.
Overall, the computational complexity of the STA MLD selection procedure for DL communications ($\beta=0$) is $\mathcal{O}(M)$.

As a result, the computational complexity of the proposed non-cooperative approach is $\mathcal{O}(k^3)$ for UL sensing ($\beta=1$) and $\mathcal{O}(M)$ for DL communications ($\beta=0$).
The lightweight computational complexity renders the proposed non-cooperative approach feasible for deployment in the AP MLD.

\section{Cooperative Approach}
\label{sec:Coop}
While the non-cooperative approach proposed in Sec. \ref{sec:Non-Coop} requires only \emph{local} information, it may suffer from a critical limitation without the cooperation among all interfaces of the AP MLD: When it is time for the $l$th interface to conduct UL sensing ($t>t^*_l$), it may experience obstruction due to an insufficient number of available STA MLDs ($|\mathcal{I}^a_l|<3$) or may reluctantly select three STA MLDs from candidates with low UL SNRs, which may eventually degrade the sensing performance for target tracking.

By properly enabling the cooperation among all interfaces, such limitation can be removed.
In this section, we propose a cooperative approach, where each interface works with the aggregate information across all interfaces upon gaining a TXOP under an EMLSR operation.
For the proposed cooperative approach, we describe its procedures of ISAC decision and STA MLD selection, and we analyze its computational complexity, focusing on its difference from the proposed non-cooperative approach.
An illustration of the proposed cooperative approach is shown in Fig. \ref{fig:illustration_C}.

\subsection{ISAC Decision}
\label{subsec:C_ISAC_decision}
The proposed cooperative approach begins with the procedure of ISAC decision, which creates a predicted state of the target and an action between sensing and communications, as illustrated in Algorithm \ref{algo:C_ISAC}.

\begin{algorithm}
\SetAlgoLined

\textbf{Input:} $\tilde{\mathbf{x}}_L',\mathcal{I}_l^a,t,t_L',t_n,t_E,N_L,\alpha$

\textbf{Initialization:} $T'=t-t_L',T^{\#}=t_E-t$

$\hat{\mathbf{x}}_L=\mathbf{F}\tilde{\mathbf{x}}_L'$

\textbf{if} $\mathcal{I}_l^a\neq\varnothing$ \textbf{and} $T^{\#}\geq\tau_{min}$

\quad $t_L^*=\alpha^{N_L+1}t_L'+(1-\alpha^{N_L+1})t_E$

\quad \textbf{if} $t\leq t_L^*-\tau_{c,min}$

\qquad $\beta=0$

\quad \textbf{else if} $t<\mbox{min}\{t_L'+\tau_{s,min},t_n-\tau_{c,min}\}$

\qquad $t_L^*=t_n$; $\beta=0$


\quad \textbf{else if} $t>\mbox{max}\{t_L^*,t_L'+\tau_{s,min}\}$

\qquad $\beta=1$

\quad \textbf{end if}

\textbf{end if}

\textbf{Output:} $\hat{\mathbf{x}}_L,\beta$

\caption{Cooperative ISAC Decision}
\label{algo:C_ISAC}
\end{algorithm}

Over $L$ interfaces, the last sensing TXOP occurred at time $t_L'$ with the updated state $\tilde{\mathbf{x}}_L'$ generated, and the next available TXOP will occur at time $t_n$.
We initialize the elapsed time from last sensing TXOP $T'=t-t_L'$ and the remaining time in current time window $T^{\#}=t_E-t$.
When the $l$th interface gains a TXOP with $\mathcal{I}^a_l$ at time $t$, the aggregate information across $L$ interfaces is considered, which facilitates a more comprehensive decision.
According to (\ref{eq:hat_x}), we compute a predicted state of the target as
\begin{equation}
    \hat{\mathbf{x}}_L=\mathbf{F}\tilde{\mathbf{x}}_L'.
    \label{eq:C_predicted_state}
\end{equation}

Likewise, an action between sensing and communications is to be determined if both conditions (\ref{eq:cond_nonempty}) and (\ref{eq:cond_time}) are fulfilled.
With the number of previous sensing TXOPs in the current time window $N_L$ over $L$ interfaces taken into account, we follow (\ref{eq:t_star}) and compute
\begin{equation}
    t_L^*=\alpha^{N_L+1}t_L'+(1-\alpha^{N_L+1})t_E.
    \label{eq:C_tl_star}
\end{equation}
In order to remove the critical limitation (from which the proposed non-cooperative approach may suffer), we aim to ensure that the $l$th interface has access to all of $M$ STA MLDs (i.e., $\mathcal{I}^a_l=\mathcal{M}$) when it conducts sensing ($\beta=1$).
Particularly, we achieve this aim by having the $l$th interface follow the three criteria below (which ultimately lead to the cooperation among all interfaces):
\begin{enumerate}
    \item The $l$th interface should conduct communications ($\beta=0$) if accomplishing a communications TXOP by time $t_L^*$ is possible ($t\leq t^*_L-\tau_{c,min}$).
    \item The $l$th interface should conduct communications ($\beta=0$) with $t^*_L=t_n$ if the last sensing TXOP is ongoing and accomplishing a communications TXOP by time $t_n$ is possible ($t<\mbox{min}\{t_L'+\tau_{s,min},t_n-\tau_{c,min}\}$).
    \item The $l$th interface should conduct sensing ($\beta=1$) after time $t_L^*$ and after the last sensing TXOP is accomplished ($t>\mbox{max}\{t_L^*,t_L'+\tau_{s,min}\}$).
\end{enumerate}
Namely, if each interface follows the three criteria above, then the cooperation among all interfaces is enabled and the limitation can be removed.

\subsection{STA MLD Selection}
\label{subsec:C_SMS}
Depending on the value of $\beta$ (with the cooperation among all interfaces enabled), the proposed cooperative approach proceeds with the procedure of STA MLD selection for UL sensing ($\beta=1$) or DL communications ($\beta=0$), as illustrated in Algorithm \ref{algo:C_SMS}.

\begin{algorithm}
\SetAlgoLined

\textbf{Input:} $\beta$

\textbf{if} $\beta=1$

\quad \textbf{Input:} $\hat{\mathbf{x}}_L,\{\xi^u_{m,l}\}_{m\in\mathcal{M}},\{(\bar{x}_m,\bar{y}_m)\}_{m\in\mathcal{M}},k$

\quad \textbf{if} $M>k$

{\fontsize{9}{6}
\qquad \textbf{Sort:} $\xi^u_{j_1,l}\geq\xi^u_{j_2,l}\geq\dots\geq\xi^u_{j_M,l},j_1,j_2,\dots,j_M\in\mathcal{M}$
}

\qquad $\mathcal{I}^c_l=\{j_1,j_2,\dots,j_k\}$

\quad \textbf{else if} $M\leq k$

\qquad $\mathcal{I}^c_l=\mathcal{M}$

\quad \textbf{end if}

\quad $\mathcal{I}_l=\underset{\mathcal{I}_l'\in[\mathcal{I}^c_l]^3}{\mbox{argmin}}\ \mbox{Tr}\{\hat{\mathbf{\Psi}}_{\mathcal{I}_l',l}^{-1}\}$

\quad \textbf{Trilateration measurement:} $\mathbf{z}$

\quad $\tilde{\mathbf{x}}_L=\hat{\mathbf{x}}_L+\mathbf{K}(\mathbf{z}-\mathbf{H}\hat{\mathbf{x}}_L)$

\textbf{Output:} $\mathcal{I}_l,\tilde{\mathbf{x}}_L$

\textbf{else if} $\beta=0$

\quad \textbf{Input:} $\{(b_m^r,b_m^x)\}_{m\in\mathcal{I}_l^a},p^*_l$

\quad \textbf{Initialization:} $\mathcal{I}_l=\varnothing,p'_l=p^*_l$

\quad \textbf{z-score normalization:} $\{b_m^r\}_{m\in\mathcal{I}_l^a}\rightarrow\{z_{s_m}\}_{m\in\mathcal{I}_l^a}$

\quad\textbf{for} $m\in\mathcal{I}_l^a$

\qquad $w_m=\mbox{exp}(-z_{s_m})$; $\psi_m=w_m\,\mbox{log}(b^x_m)/b^x_m$

\quad\textbf{end for}

{\fontsize{9}{6}
\quad\textbf{Sort:} $\psi_{j_1'}\geq \psi_{j_2'}\geq\dots\geq \psi_{j_{|\mathcal{I}_l^a|}'},j_1',j_2',\dots,j_{|\mathcal{I}_l^a|}'\in\mathcal{I}_l^a$
}

\quad\textbf{for} $q=1:|\mathcal{I}_l^a|$

\qquad $\mathcal{I}_l\leftarrow\mathcal{I}_l\cup j_q'$; $p'_l\leftarrow p'_l-b_{j_q'}^x$

\qquad\textbf{if} $p'_l<0$

\quad\qquad \textbf{break}

\qquad\textbf{end if}

\quad\textbf{end for}

\quad\textbf{Output:} $\mathcal{I}_l$

\textbf{end if}

\caption{Cooperative STA MLD Selection}
\label{algo:C_SMS}
\end{algorithm}

If an action of sensing is determined ($\beta=1$), then the $l$th interface has access to all of $M$ STA MLDs (i.e., $\mathcal{I}^a_l=\mathcal{M}$), given the enabled cooperation among all interfaces.
With access to all of $M$ STA MLDs, we are required to select three STA MLDs of indices $\mathcal{I}_l\in[\mathcal{M}]^3$ for trilateration in UL sensing.

For the index set of candidate STA MLDs $\mathcal{I}^c_l\subseteq\mathcal{M}$ following (\ref{eq:candidate}) based on the developed $k$-candidates strategy, we nominate $k$ candidates with highest UL SNRs as $\mathcal{I}^c_l$ if $M>k$, while all of $M$ STA MLDs are nominated as candidates, i.e., $\mathcal{I}^c_l=\mathcal{M}$, if $M\leq k$.
Then, we select three STA MLDs from the candidates in $\mathcal{I}^c_l$ with minimum predicted CRLB of trilateration estimate of indices $\mathcal{I}_l$ as (\ref{eq:sensing_STA_MLD}).
Leveraging the predicted state $\hat{\mathbf{x}}_L$ in (\ref{eq:C_predicted_state}) and the trilateration measurement $\mathbf{z}$ obtained with the three STA MLDs of indices $\mathcal{I}_l$ in (\ref{eq:sensing_STA_MLD}), we compute an updated state of the target, according to (\ref{eq:tilde_x}), as
\begin{equation}
    \tilde{\mathbf{x}}_L=\hat{\mathbf{x}}_L+\mathbf{K}(\mathbf{z}-\mathbf{H}\hat{\mathbf{x}}_L).
    \label{eq:C_updated_state}
\end{equation}

If an action of communications is determined ($\beta=0$), then we are required to select some STA MLDs of indices $\mathcal{I}_l\subseteq\mathcal{I}^a_l$ for data transmission in DL communications, which should be accomplished by time $t_L^*$ according to the first criterion stated in Sec. \ref{subsec:C_ISAC_decision}.

To start with, we initialize (\ref{init:set}) and
\begin{equation}
    p'_l=p^*_l,
\end{equation}
where $p^*_l$ is the maximum number of bytes to be transmitted DL over the $l$th channel by time $t_L^*$.
Then, we normalize the number of bytes that have been received DL $\{b^r_m\}_{m\in\mathcal{I}^a_l}$ into z-score $\{z_{s_m}\}_{m\in\mathcal{I}^a_l}$, and we obtain the weight $w_m$ (dependent on z-score $z_{s_m}$) with (\ref{eq:weight}) and the average weighted utility per byte $\psi_m$ with (\ref{eq:avg_weighted_utility}) for the $m$th STA MLD.
Subsequently, we sort the average weighted utility per byte of the STA MLDs in descending order as $\psi_{j_1'}\geq \psi_{j_2'}\geq\dots\geq \psi_{j_{|\mathcal{I}_l^a|}'}$, where $j_1',j_2',\dots,j_{|\mathcal{I}_l^a|}'\in\mathcal{I}_l^a$ is the order of indices of STA MLDs to be addressed.
In the $q$th iteration, the $j_q'$th STA MLD is addressed with its index $j_q'$ added to $\mathcal{I}_l$ and its number of bytes to be transmitted DL $b^x_{j_q'}$ subtracted from $p_l'$.
Once $p_l'$ goes below zero, the STA MLD selection stops with $\mathcal{I}_l$ obtained.

\subsection{Computational Complexity}
With its procedures of ISAC decision (Algorithm \ref{algo:C_ISAC}) and STA MLD selection (Algorithm \ref{algo:C_SMS}) as described in Secs. \ref{subsec:C_ISAC_decision} and \ref{subsec:C_SMS}, respectively, the proposed cooperative approach has each interface work with the aggregate information across all interfaces, among which the cooperation is enabled, under the AP MLD. Below, we analyze the computational complexity of the proposed cooperative approach in terms of the number of multiplications/divisions involved.

First, we look into the ISAC decision procedure (Algorithm \ref{algo:C_ISAC}).
The computation of predicted state in (\ref{eq:C_predicted_state}) involves two multiplications (with two non-binary entries of $\mathbf{F}$ of value $T'>0$), and the computation of $t_L^*$ in (\ref{eq:C_tl_star}) involves at most three multiplications (with the update of $\alpha^{N_L+1}$ involving at most one multiplication with $\alpha$ by maintaining the value of $N_L$ and $\alpha^{N_L+1}$ throughout the TXOPs over $L$ interfaces).
Overall, the computational complexity of the ISAC decision procedure (Algorithm \ref{algo:C_ISAC}) is $\mathcal{O}(1)$.

Next, we look into the STA MLD selection procedure (Algorithm \ref{algo:C_SMS}).
The first case corresponds to UL sensing ($\beta=1$).
Consider any $\mathcal{I}_l=\{i_{l,1},i_{l,2},i_{l,3}\}\in[\mathcal{M}]^3$.
Note that the computation of distance $\hat{d}_{i_{l,j}}$ between predicted position of target in (\ref{eq:C_predicted_state}) and position of the $i_{l,j}$th STA MLD involves $\mathcal{O}(1)$ multiplications, $j=1,2,3$.
Then, the computation of  $\hat{\mathbf{\Gamma}}_{\mathcal{I}_l,l}$ in (\ref{eq:Gamma_symbol}) involves six divisions, and the computation of $\mathbf{D}_{\mathcal{I}_l,l}$ in (\ref{eq:D_symbol}) involves three multiplications (by maintaining the value of $\omega_l^2/\mu$).
While the computation of $\hat{\mathbf{\Psi}}_{\mathcal{I}_l,l}$ in (\ref{eq:hat_Psi_symbol}) involves fifteen multiplications (with each $j$ index involving five multiplications arising from $\gamma_{1,j}\rho_j$ and $\gamma_{2,j}\rho_j$ along with $\gamma_{1,j}^2\rho_j$, $\gamma_{2,j}^2\rho_j$, and $\gamma_{1,j}\gamma_{2,j}\rho_j$), the computation of inverse $\hat{\mathbf{\Psi}}_{\mathcal{I}_l,l}^{-1}$ involves two multiplications and one division.
Therefore, the computational complexity of deriving predicted CRLB of trilateration estimate in (\ref{eq:hat_CRLB_tri}) given $\mathcal{I}_l$ is $\mathcal{O}(1)$.
Accordingly, the computational complexity of the selection of three STA MLDs (from up to $k$ candidates) with minimum predicted CRLB of trilateration estimate in (\ref{eq:sensing_STA_MLD}) is $\mathcal{O}(k^3)$.
The computation of updated state in (\ref{eq:C_updated_state}) involves eight multiplications.
Overall, the computational complexity of the STA MLD selection procedure for UL sensing ($\beta=1$) is $\mathcal{O}(k^3)$.
The second case corresponds to DL communications ($\beta=0$).
Note that the z-score normalization involves $\mathcal{O}(M)$ multiplications/divisions.
For the $\mathcal{O}(M)$ STA MLDs belonging to $\mathcal{I}^a_l$, the computation of their weight in (\ref{eq:weight}) and their average weighted utility per byte in (\ref{eq:avg_weighted_utility}) involves $\mathcal{O}(M)$ multiplications/divisions.
Overall, the computational complexity of the STA MLD selection procedure for DL communications ($\beta=0$) is $\mathcal{O}(M)$.

As a result, the computational complexity of the proposed cooperative approach is $\mathcal{O}(k^3)$ for UL sensing ($\beta=1$) and $\mathcal{O}(M)$ for DL communications ($\beta=0$), comparable to that of the proposed non-cooperative approach.
Similarly, the lightweight computational complexity makes the proposed cooperative approach a feasible scheme to be deployed in the AP MLD.

\section{Simulation}
\label{sec:simulation}
In this section, we evaluate the performance of the proposed non-cooperative and cooperative approaches with MATLAB simulation of a Wi-Fi network featuring both EMLSR operation and ISAC functionality.
With respect to the proposed non-cooperative and cooperative approaches, we investigate the choice of $k$ value (from the developed $k$-candidates strategy), examine the effect of number of STA MLDs, and compare the original scheme with the following random STA MLD selection schemes (reduced from the original scheme):
\begin{itemize}
    \item \textbf{Random STA MLD selection for UL sensing (RSMS-S)}: The STA MLD index set $\mathcal{I}_l$ is randomly selected from $[\mathcal{I}^a_l]^3$ in non-cooperative approach or from $[\mathcal{M}]^3$ in cooperative approach for UL sensing.
    \item \textbf{Random STA MLD selection for DL communications (RSMS-C)}: The STA MLD index set $\mathcal{I}_l$ is randomly selected as a subset of $\mathcal{I}^a_l$ for DL communications.
    \item \textbf{Random STA MLD selection for UL sensing and DL communications (RSMS-SC)}: Both RSMS-S and RSMS-C are enabled.
\end{itemize}

\subsection{Parameter Settings}
The Wi-Fi network is composed of an AP MLD and $M$ STA MLDs which are randomly located with x and y coordinates uniformly chosen from $[-10,10]$ m along with a moving target with initial position at origin $(0,0)$ and initial velocity of 1 m/s in a random direction on a 2D area.
Each MLD hosts $L=3$ interfaces of carrier frequency 2.437, 5.250, and 6.295 GHz with respective channels of channel bandwidth 40, 80, and 160 MHz.
We compute the byte upper bound $\{p_l\}_{l=1}^L$ with the Shannon-Hartley theorem \cite{shannon1948}.
For brevity, the core Wi-Fi network parameter settings are summarized in Table \ref{tab_sims}.

\begin{table}[h]
    \centering
    \caption{Wi-Fi Network Parameter Settings}
    \label{tab_sims}
    \begin{tabular}{c|c}
    \hline
    Parameter     & Value\\
    \hline\hline
    AP MLD and STA MLD position & Random\\
    \hline
    Target initial position & $(0,0)$\\
    \hline
    Target initial velocity & 1 m/s in random direction\\
    \hline
    \# interfaces $L$  & 3\\
    \hline
     Carrier frequency    & 2.437, 5.250, 6.295 GHz\\
     \hline
     Channel bandwidth    & 40, 80, 160 MHz\\
     \hline
     Time window duration $\tau_w$  & 10.24 ms\\
     \hline
     \# time windows  & 200\\
     \hline
     DL data rate for an STA MLD & 20 Mbps\\
     \hline
     ($\tau_{SIFS}$, $\tau_{TF}$, $\tau_{CTS}/\tau_{ACK}$, $\tau_{NDP}$) & (16, 10.8, 4.6, 44+8$\rho\eta$) $\mu\mbox{s}$\\
     \hline
     \# EHT-LTF symbols $\rho$ & 4\\
     \hline
     \# EHT-LTF repetitions $\eta$ & 4\\
     \hline
     Process noise intensity $g_s$  & 0.1\\
     \hline
     Channel model  & IEEE 802.11be indoor\\
     \hline
     AP MLD interface Tx power  & 43 dBm\\
     \hline
     STA MLD interface Tx power & 23 dBm\\
     \hline
     Multiple-input multiple-output (MIMO)  & $4\times 2$\\
     \hline
    \end{tabular}
\end{table}

\subsection{Choice of $k$ value}
\begin{figure}[htbp]%
\centering
\subfigure[MSE between target position $(x,y)$ and predicted target position $(\hat{x},\hat{y})$]{%
\label{fig:k_MSE}%
\includegraphics[width=8.5cm]{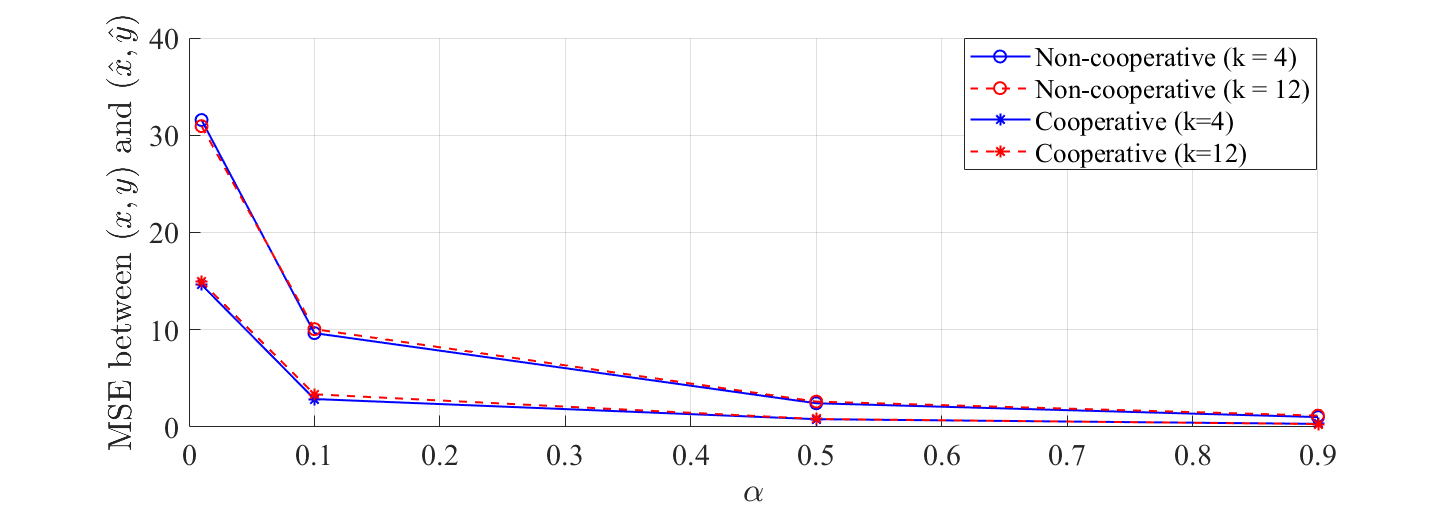}}%
\qquad
\subfigure[Throughput]{%
\label{fig:k_throughput}%
\includegraphics[width=8.5cm]{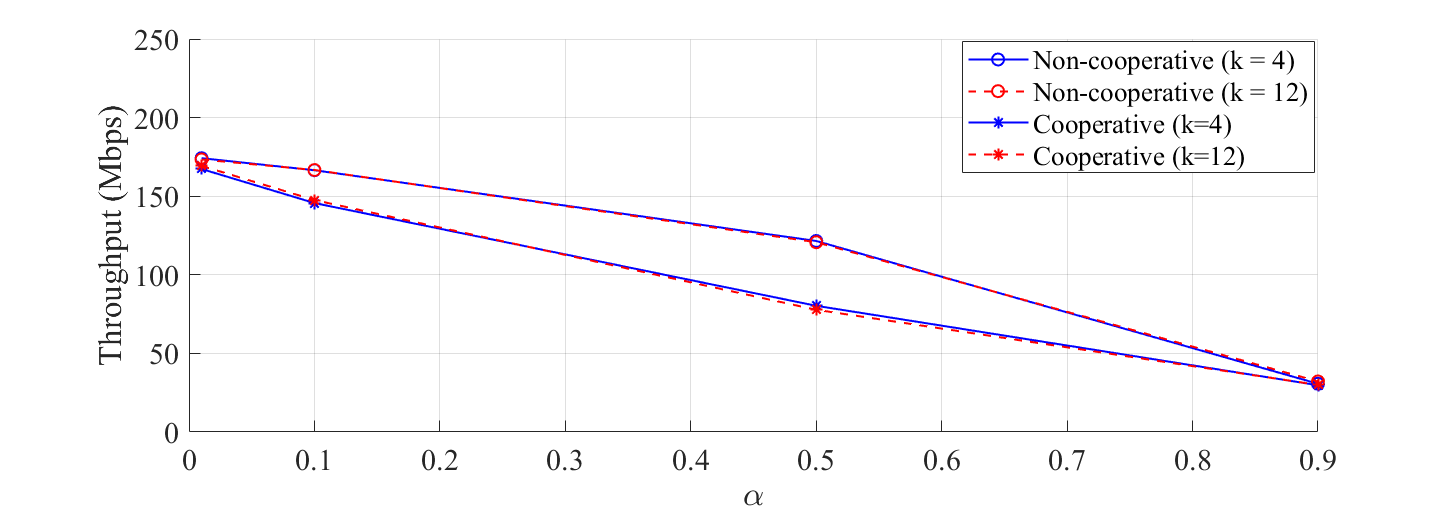}}%
\caption{MSE and throughput under different values $k=\{4,12\}$ with number of STA MLDs $M = 12$}
\label{fig:approach_k}
\end{figure}

First, we investigate the choice of $k$ value from the perspectives of sensing (in terms of MSE between target position $(x,y)$ and predicted target position $(\hat{x},\hat{y})$) and communications (in terms of throughput) with number of STA MLDs $M=12$ over $\alpha=\{0.01,0.1,0.5,0.9\}$ across $k=\{4,12\}$.
The results are shown in Fig. \ref{fig:approach_k}, where the MSE and the throughput are demonstrated in Figs. \ref{fig:k_MSE} and \ref{fig:k_throughput}, respectively.

The results of $k=4$ are very close to those of $k=12$, which implies that a modest $k$ value such as $k=4$ suffices for the $k$-candidates strategy.
Therefore, we adopt $k=4$ for subsequent evaluations.
As $\alpha$ increases (which favors sensing more), there is a reduction in MSE (due to more sensing) and in throughput (due to less communications).
Besides, the comparison of non-cooperative and cooperative approaches reveals a tradeoff between sensing and communications under fixed resources.
While the cooperative approach improves sensing with lower MSE at the cost of lower throughput, the non-cooperative approach achieves higher throughput with the sacrifice of higher MSE.

\subsection{Effect of Number of STA MLDs}
\begin{figure}[ht]
\centering
\includegraphics[width=8.5cm]{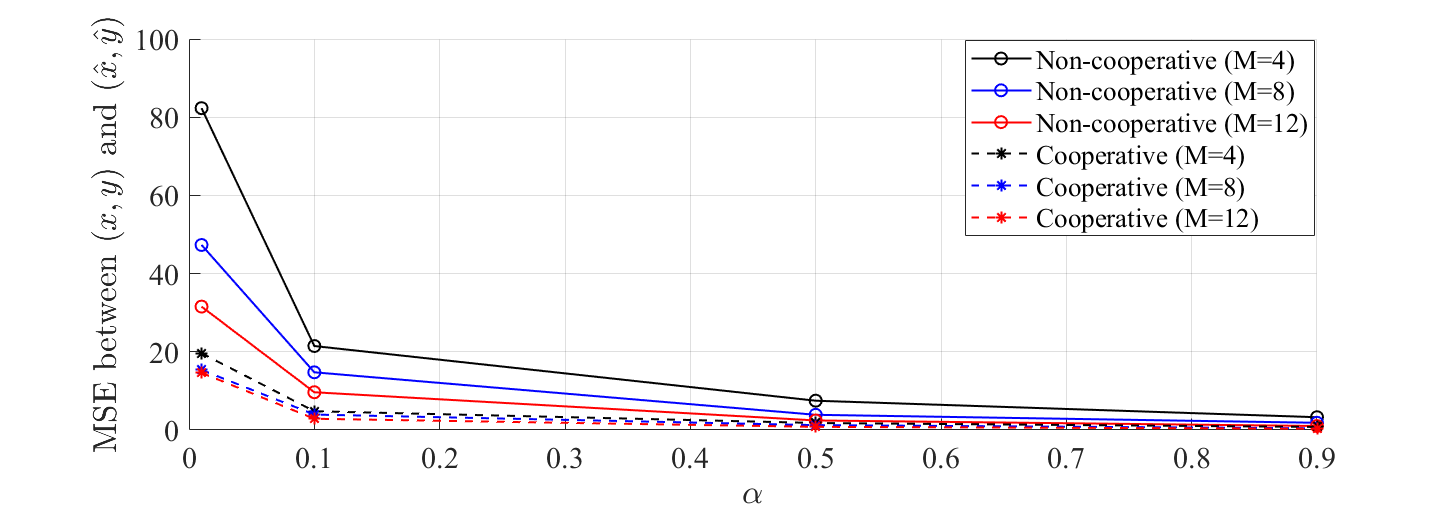}
\caption{MSE between target position $(x,y)$ and predicted target position $(\hat{x},\hat{y})$ under different number of STA MLDs $M=\{4,8,12\}$ with $k=4$}
\label{fig:STA_num_MSE}
\end{figure}

Second, we examine the effect of number of STA MLDs on sensing (in terms of MSE between target position $(x,y)$ and predicted target position $(\hat{x},\hat{y})$) with $k=4$ over $\alpha=\{0.01,0.1,0.5,0.9\}$ across $M=\{4,8,12\}$.
The MSE results are shown in Fig. \ref{fig:STA_num_MSE}.

Similarly, the non-cooperative approach leads to higher MSE, while the cooperative approach leads to lower MSE.
Also, an increase in $\alpha$ reduces both MSE and throughput.
As the number of STA MLDs increases, there is a reduction in MSE (due to the smaller predicted CRLB of trilateration estimate achieved by the three STA MLDs selected from a larger pool).

\subsection{Comparison of Different Schemes}
\begin{figure}[htbp]%
\centering
\subfigure[MSE between target position $(x,y)$ and predicted target position $(\hat{x},\hat{y})$]{%
\label{fig:rand_STA_MSE}%
\includegraphics[width=8.5cm]{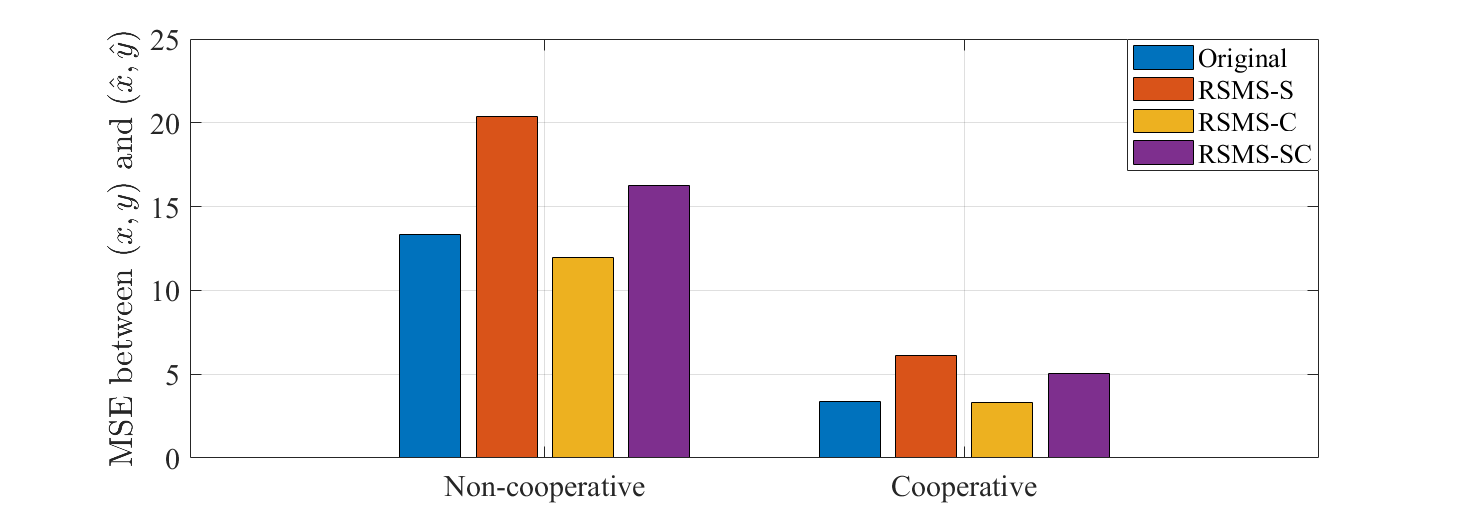}}%
\qquad
\subfigure[Throughput]{%
\label{fig:rand_STA_throughput}%
\includegraphics[width=8.5cm]{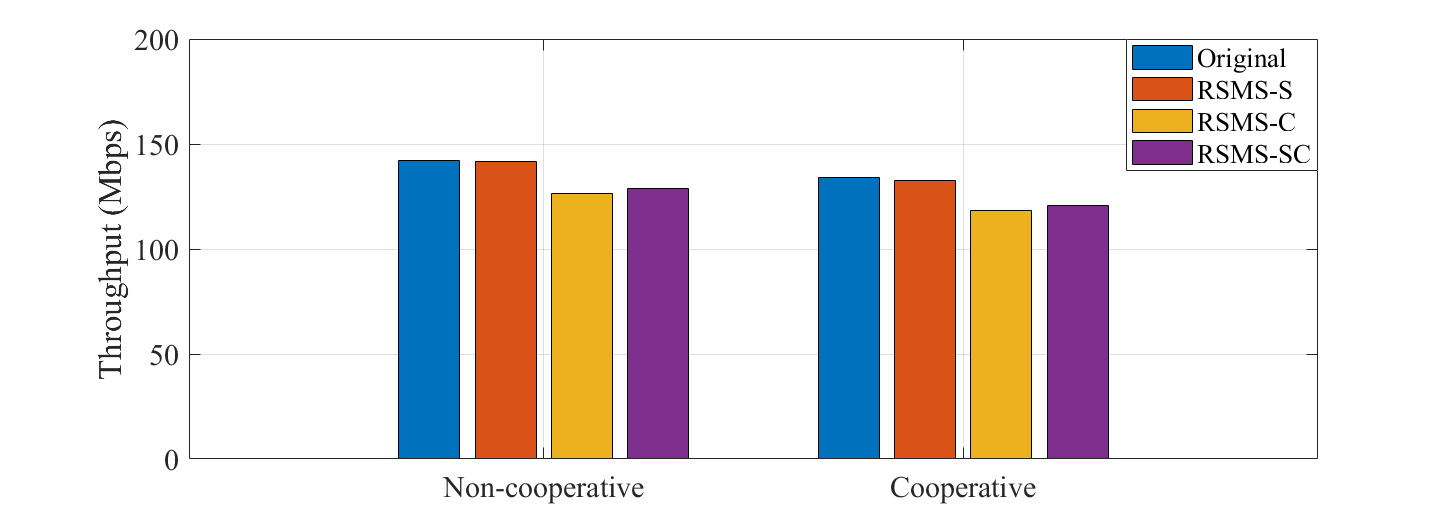}}%
\qquad
\subfigure[Jain’s fairness index]{%
\label{fig:rand_STA_fairness}%
\includegraphics[width=8.5cm]{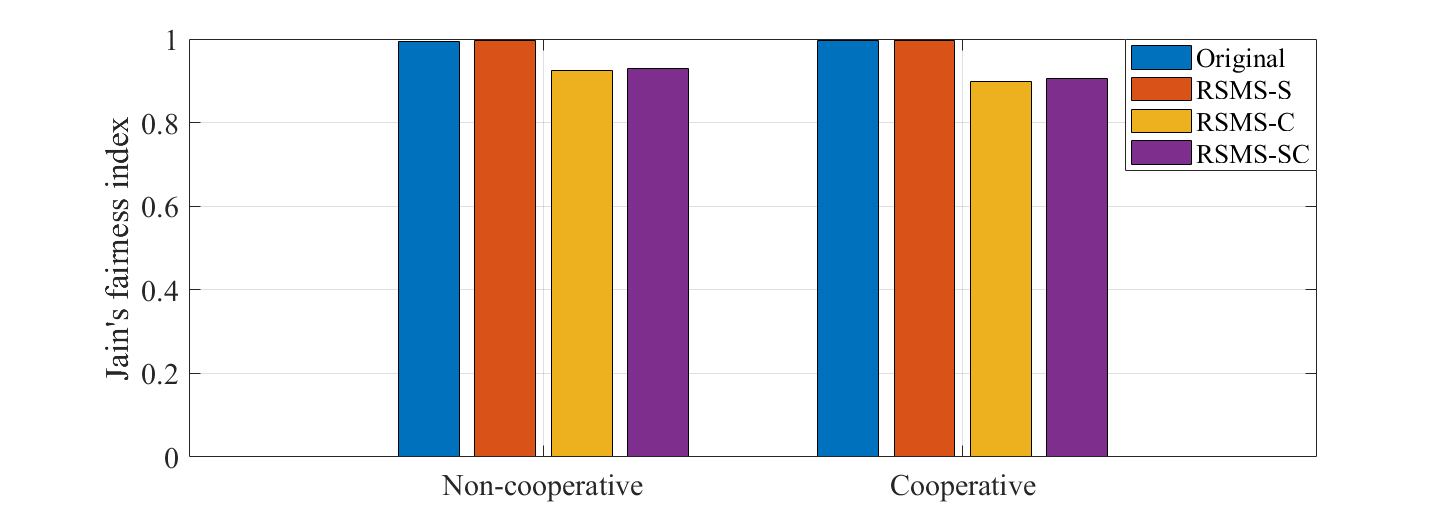}}%
\caption{MSE, throughput, and Jain's fairness index under different schemes with number of STA MLDs $M = 8$, $k=4$, and $\alpha=0.5$}
\label{fig:approach_rand}
\end{figure}

Third, we compare different schemes (Original, RSMS-S, RSMS-C, and RSMS-SC) according to their performance of sensing (in terms of MSE between target position $(x,y)$ and predicted target position $(\hat{x},\hat{y})$) and communications (in terms of throughput and Jain's fairness index \cite{jain1984}) with $k=4$, $\alpha=0.5$, and $M=8$.
The results are shown in Fig. \ref{fig:approach_rand}, where Fig. \ref{fig:rand_STA_MSE} demonstrates the MSE, Fig. \ref{fig:rand_STA_throughput} the throughput, and Fig. \ref{fig:rand_STA_fairness} the Jain's fairness index, respectively.

While the non-cooperative and cooperative approaches exhibit a similar trend in terms of MSE and throughput (i.e., MSE and throughput are higher with the non-cooperative approach and are lower with the cooperative approach), they achieve comparable performance in terms of Jain's fairness index, which validates the efficacy of the developed weighted proportional fairness-aware heuristic strategy for DL communications.
Between Original and RSMS-S, Original demonstrates superiority in terms of MSE (thanks to its careful STA MLD selection for UL senisng) and demonstrates a comparable performance in terms of throughput and Jain's fairness index (due to the same STA MLD selection procedure for DL communications).
Between Original and RSMS-C, Original demonstrates superiority in terms of throughput and Jain's fairness index (thanks to its careful STA MLD selection for DL communications) and demonstrates a comparable performance in terms of MSE (due to the same STA MLD selection procedure for UL sensing).
Between Original and RSMS-SC, Original demonstrates superiority in terms of MSE, throughput, and Jain's fairness index (thanks to its careful STA MLD selection for UL sensing and for DL communications).

\section{Conclusion}
\label{sec:conclusion}
In this paper, we address target tracking with ISAC using EMLSR in IEEE 802.11 WLANs.
Within our established scenario, the AP MLD is required to make an ISAC decision and an STA MLD selection when its interface gains a TXOP.
With the key design principles, we adopt the Kalman filter to track the state of target and develop a time-based strategy to determine an action between sensing and communications in ISAC decision, and we take the predicted CRLB of trilateration estimate along with the developed $k$-candidates strategy for UL sensing and develop a weighted proportional fairness-aware heuristic strategy for DL communications in STA MLD selection.
Based on the design principles of ISAC decision and STA MLD selection, we propose novel non-cooperative and cooperative approaches. 
Each interface utilizes its own information with the proposed non-cooperative approach and utilizes the aggregate information across all interfaces with the proposed cooperative approach.
According to simulation results, the proposed non-cooperative and cooperative approaches reveal a tradeoff between sensing and communications and demonstrate their superiority in terms of sensing and communications.
Future work for this scenario includes multi-target tracking and the combination with multi-AP coordination proposed in IEEE 802.11bn (envisioned as Wi-Fi 8).



%

\appendices
\section{Proof of Theorem 1}
\label{pf:thm1}
Consider any $\mathcal{I}_l=\{i_{l,1},i_{l,2},i_{l,3}\}\in[\mathcal{I}^a_l]^3$ with its corresponding $\hat{\mathbf{\Psi}}_{\mathcal{I}_l,l}=\hat{\mathbf{\Gamma}}_{\mathcal{I}_l,l}\mathbf{D}_{\mathcal{I}_l,l}\hat{\mathbf{\Gamma}}_{\mathcal{I}_l,l}^T\in\mathbf{S}^2_{++}$.
Denote the two positive eigenvalues of $\hat{\mathbf{\Psi}}_{\mathcal{I}_l,l}$ as $\lambda_1>0$ and $\lambda_2>0$, respectively.
Express
\begin{equation}
\hat{\mathbf{\Gamma}}_{\mathcal{I}_l,l}=\begin{bmatrix}
    \gamma_{1,1} & \gamma_{1,2} & \gamma_{1,3} \\
    \gamma_{2,1} & \gamma_{2,2} & \gamma_{2,3}
\end{bmatrix},
\label{eq:Gamma_symbol}
\end{equation}
where $\gamma_{1,j}=(\hat{x}-\bar{x}_{i_{l,j}})/\hat{d}_{i_{l,j}}$ and $\gamma_{2,j}=(\hat{y}-\bar{y}_{i_{l,j}})/\hat{d}_{i_{l,j}}$ with
\begin{equation}
    \gamma_{1,j}^2+\gamma_{2,j}^2=1,
    \label{eq:two_square_equal_unity}
\end{equation} 
and
\begin{equation}
    \mathbf{D}_{\mathcal{I}_l,l}=\mbox{diag}([\rho_1\,\rho_2\,\rho_3]^T),
    \label{eq:D_symbol}
\end{equation}
where $\rho_j=C_{r_{i_{l,j},l}}^{-1}=(\omega_l^2\xi^u_{i_{l,j},l})/\mu,j=1,2,3$.

For any square matrix, its trace is equal to the sum of its eigenvalues.
Therefore, the trace of $\hat{\mathbf{\Psi}}_{\mathcal{I}_l,l}$ can be obtained as
\begin{equation}
    \mbox{Tr}\{\hat{\mathbf{\Psi}}_{\mathcal{I}_l,l}\}=\lambda_1+\lambda_2.
    \label{eq:trace_eigenvalue}
\end{equation}
Since the eigenvalues of a diagonal matrix are equal to its diagonal entries, the trace of $\mathbf{D}_{\mathcal{I}_l,l}$ can be also obtained as
\begin{equation}
    \mbox{Tr}\{\mathbf{D}_{\mathcal{I}_l,l}\}=\sum_{j=1}^3\rho_j.
    \label{eq:trace_diagonal}
\end{equation}
Substituting (\ref{eq:Gamma_symbol}) and (\ref{eq:D_symbol}) into $\hat{\mathbf{\Psi}}_{\mathcal{I}_l,l}$, we derive
\begin{equation}
\hat{\mathbf{\Psi}}_{\mathcal{I}_l,l}=\begin{bmatrix}
    \sum_{j=1}^3\gamma_{1,j}^2\rho_j & \sum_{j=1}^3\gamma_{1,j}\gamma_{2,j}\rho_j \\
    \sum_{j=1}^3\gamma_{1,j}\gamma_{2,j}\rho_j & \sum_{j=1}^3\gamma_{2,j}^2\rho_j
\end{bmatrix}.
\label{eq:hat_Psi_symbol}
\end{equation}
Then, the trace of $\hat{\mathbf{\Psi}}_{\mathcal{I}_l,l}$ can be alternatively obtained as
\begin{equation}
    \mbox{Tr}\{\hat{\mathbf{\Psi}}_{\mathcal{I}_l,l}\}=\sum_{j=1}^3(\gamma_{1,j}^2+\gamma_{2,j}^2)\rho_j=\sum_{j=1}^3\rho_j=\mbox{Tr}\{\mathbf{D}_{\mathcal{I}_l,l}\},
    \label{eq:trace_Psi_D}
\end{equation}
where the second and third equalities hold according to (\ref{eq:two_square_equal_unity}) and (\ref{eq:trace_diagonal}), respectively.
Incorporating (\ref{eq:trace_eigenvalue}) and (\ref{eq:trace_Psi_D}), we obtain
\begin{equation}
    \mbox{Tr}\{\hat{\mathbf{\Psi}}_{\mathcal{I}_l,l}\}=\lambda_1+\lambda_2=\sum_{j=1}^3\rho_j.
    \label{eq:lambda_rho}
\end{equation}

For any invertible matrix, the trace of its inverse is equal to the sum of reciprocals of its eigenvalues.
Therefore, the trace of $\hat{\mathbf{\Psi}}_{\mathcal{I}_l,l}^{-1}$ can be obtained as
\begin{equation}
    \mbox{Tr}\{\hat{\mathbf{\Psi}}_{\mathcal{I}_l,l}^{-1}\}=1/\lambda_1+1/\lambda_2=(\lambda_1+\lambda_2)/(\lambda_1\lambda_2).
    \label{eq:trace_inverse}
\end{equation}
Applying the AM-GM inequality to the two positive eigenvalues $\lambda_1,\lambda_2>0$, we obtain
\begin{equation}
    (\lambda_1+\lambda_2)/2\geq\sqrt{\lambda_1\lambda_2},
\end{equation}
which is equivalent to
\begin{equation}
    \lambda_1\lambda_2\leq(\lambda_1+\lambda_2)^2/4.
    \label{ineq:lambda}
\end{equation}
Substituting (\ref{ineq:lambda}) into (\ref{eq:trace_inverse}), we derive the lower bound of $\mbox{Tr}\{\hat{\mathbf{\Psi}}_{\mathcal{I}_l,l}^{-1}\}$ as
\begin{equation}
    \mbox{Tr}\{\hat{\mathbf{\Psi}}_{\mathcal{I}_l,l}^{-1}\}\geq 4/(\lambda_1+\lambda_2)=4/\sum_{j=1}^3\rho_j,
    \label{ineq:first_lower_bound}
\end{equation}
where the second equality holds according to (\ref{eq:lambda_rho}).

According to (\ref{ineq:first_lower_bound}), for all $\hat{\mathbf{\Psi}}\in\{\hat{\mathbf{\Psi}}_{\mathcal{I}_l,l}:\mathcal{I}_l\in[\mathcal{I}_l^a]^3\}$, the lower bound of $\mbox{Tr}\{\hat{\mathbf{\Psi}}^{-1}\}$ can be obtained as
\begin{equation}
    \mbox{Tr}\{\hat{\mathbf{\Psi}}^{-1}\}\geq\underset{\mathcal{I}_l\in[\mathcal{I}^a_l]^3}{\mbox{min}}4/\sum_{j=1}^3\rho_j,
\end{equation}
which is equivalent to
\begin{equation}
    \mbox{Tr}\{\hat{\mathbf{\Psi}}^{-1}\}\geq4/\underset{\mathcal{I}_l\in[\mathcal{I}^a_l]^3}{\mbox{max}}\sum_{j=1}^3\rho_j=4\mu/(\omega_l^2\xi_l^*),
\end{equation}
where $\xi_l^*=\underset{\mathcal{I}_l=\{i_{l,1},i_{l,2},i_{l,3}\}\in[\mathcal{I}^a_l]^3}{\mbox{max}}\sum_{j=1}^3 \xi^u_{i_{l,j},l}$.



\ifCLASSOPTIONcaptionsoff
  \newpage
\fi



%


\bibliographystyle{IEEEtran}
\bibliography{IEEEabrv,waveform}

%







\end{document}

%% file: newcommands.tex
\newcommand{\cT}{{\cal T}}

\newenvironment{mat}[1]{\left[\begin{array}{#1}}{\end{array}\right]}

\renewcommand{\tt}[2]{{\cT_{#1}^{#2}}}

\newcommand{\ignore}[1]{}